\documentclass[superscriptaddress,preprint,aps,amsmath,amssymb,longbibliography,...]{revtex4-1}%

\usepackage{lineno}
\usepackage[caption=false]{subfig} 
\usepackage{color}
\usepackage{graphicx}
\usepackage{subfig}
\usepackage{ulem}
\usepackage{soul}
\usepackage{layouts}
\usepackage{xcolor}
\usepackage{diagbox}	
\usepackage{multirow}
\usepackage {vcell}
\usepackage{floatrow}
\usepackage{natbib}
\usepackage{amsmath}
\usepackage{commath}
\usepackage{enumitem}
\graphicspath{{./figures_3.0/}}

\newcommand*\diff{\mathop{}\!\mathrm{d}}	

\showboxdepth=\maxdimen
\showboxbreadth=\maxdimen

\setstcolor{red}

\begin{document} 

\floatsetup[figure]{subcapbesideposition=top}
\floatsetup[subfigure]{subcapbesideposition=top}

\title{Spectral POD analysis of the turbulent wake of a disk at $\Rey = 50,000$}

\author{S. Nidhan}
\affiliation{Department of Mechanical and Aerospace Engineering, University of California San Diego, La Jolla, CA 92093, USA}

\author{K. Chongsiripinyo}
\affiliation{Department of Mechanical Engineering, Chulalongkorn University, Bangkok, Thailand 10330}

\author{O.T. Schmidt}
\affiliation{Department of Mechanical and Aerospace Engineering, University of California San Diego, La Jolla, CA 92093, USA}

\author{S. Sarkar}
\affiliation{Department of Mechanical and Aerospace Engineering, University of California San Diego, La Jolla, CA 92093, USA}

\newcommand\Rey{\mbox{\textit{Re}}}  
\newcommand\Pran{\mbox{\textit{Pr}}} 
\newcommand\Fro{\mbox{\textit{Fr}}}  
\newcommand\Str{\mbox{\textit{St}}}  
\newcommand{\appropto}{\mathrel{\vcenter{
  \offinterlineskip\halign{\hfil$##$\cr
    \propto\cr\noalign{\kern2pt}\sim\cr\noalign{\kern-2pt}}}}}

\begin{abstract}

The coherent structures in the turbulent wake of a disk at a moderately high Reynolds number ($\Rey$) of $50,000$ are examined  using spectral proper orthogonal decomposition (SPOD) which considers all three velocity components in a  numerical database. The  SPOD eigenvalues at a given streamwise ($x$) location are functions of azimuthal wavenumber ($m$), frequency ($\Str$), and SPOD index ($n$). By $x/D =10$, two specific modes dominate the fluctuation energy: (i) the vortex shedding (VS) mode with $m=1, \Str =0.135, n=1$, and (ii) the double helix (DH) mode with $m=2, \Str \rightarrow 0, n=1$. The VS mode is more energetic than the DH mode in the near wake but, in the far wake, it is the DH mode which is  dominant. The DH mode, when scaled with local turbulent velocity and length scales, shows self-similarity in eigenvalues and eigenmodes  while the VS mode, which is a global mode, does not exhibit strict self-similarity. Modes $m = 0$, 3 and 4, although subdominant, also make a significant net contribution to the fluctuation energy, and their eigenspectra are evaluated. The reconstruction of TKE and Reynolds shear stress, $\langle u'_{x} u'_{r} \rangle$,  is evaluated by varying  $(m,\Str,n)$ combinations. Higher SPOD modes contribute significantly to the TKE, especially near the centerline. In contrast, reconstruction of  $\langle u'_{x}u'_{r}\rangle $ requires far fewer modes:  $|m| \leq 4 $,  $|\Str| \leq 1$ and $n \leq 3$. Among azimuthal modes, $m=1$ and $2$ are the leading contributors to both TKE and $\langle u'_{x}u'_{r} \rangle $. While $m=1$ captures the slope of the shear-stress profile near the centerline,  $m=2$ is important to capture $\langle u'_{x}u'_{r} \rangle $ at and near its peak.  SPOD is also performed in the  vicinity of the disk to describe the modal transition to the principal contributors in the   wake. The leading SPOD modes shows  a high-frequency  shear-layer peak close to the disk and the vortex shedding mode commences its initial dominance of the wake at the end of the recirculation region. 

\end{abstract}

\maketitle
	

\section{Introduction}

%
%
%
%
%
%
%
%
%


The turbulent wake is a widely prevalent class of free shear flows that  occurs whenever a flow encounters an obstacle in its path or, equivalently, an obstacle moves in a surrounding fluid. Like other types of free shear flows,  turbulent wakes are assumed to evolve self-similarly far away from their generators in classical analysis \cite{tennekes1972first}. Often, wakes contain large-scale anisotropic coherent structures \cite{taneda1978visual, cannon1993observations} which feed on the energy of the mean flow, and in turn modify the statistical descriptors of the flow. The characteristics of these coherent structures are strongly influenced by the geometry of wake generators and by boundary conditions. Hence, the study of turbulent wakes has revolved around two major themes: (i) discerning the scaling laws of statistical mean and turbulent quantities, and (ii) extracting and analyzing coherent structures.

Townsend \cite{townsend1976structure} hypothesized that free shear flows forget their initial conditions and eventually  asymptote towards a form that is both self-similar and  universal. However, the study of Bevilaqua and Lykoudis \cite{bevilaqua_turbulence_1978} showed that initial conditions significantly influence the subsequent evolution of turbulent wakes. They compared the wakes of a sphere and a porous disk, both of which produced the same drag, and found that, although both wakes exhibited a self-similar regime with the same power laws, their spread rates  were different. They attributed this difference to the distinct nature of coherent structures in these flows, thus pointing towards a link between the evolution of flow statistics and the nature of coherent structures. Decades later, Redford et al. \cite{redford_universality_2012} simulated a temporally evolving wake with two different types of initial conditions: (i) an array of vortex rings, and (ii) small-amplitude broadband velocity fluctuations. Both initial conditions led to a self-similarly evolving wake with the classical $U_d \sim x^{-2/3}$ decay of the wake deficit velocity ($U_d$) but the magnitude of spread rates were different. Furthermore, it was only after a very long time that the spread rates converged to a common value and a universal self-similar form with  $U_d \sim x^{-2/3}$  was achieved. 
Thus, it is possible that the imprint of coherent structures generated due to initial and boundary conditions persist for a long time (or distance) in the wake. A detailed understanding of the statistical behavior of low-dimensional coherent structures  is hence crucial to  construct a complete picture of the wake evolution. 

Early efforts to objectively study coherent structures in the context of wakes utilized laboratory experiments. Fuchs et al. \cite{fuchs_large-scale_1979} pioneered the use of two-point cross spectral analysis to investigate the coherent structures in axisymmetric shear flows. They showed the dominance of specific azimuthal modes,  $m=1$ and $m=2$,  in the wake of a disk. The azimuthal  $m=1$ mode dominated at the vortex shedding frequency of  the wake $\Str = Uf/D =  0.135$ while the $m=2$ mode peaked at a very low frequency of $\Str \approx 0.005$ in their experiments. However, the analysis was conducted at only two near-body stations,  $x/D = 3$ and $x/D = 9$, and was limited to  the fluctuating streamwise velocity ($u'_x$) and pressure ($p'$). Around the same time, flow visualizations of Taneda \cite{taneda1978visual} revealed a wavelike structure in sphere wakes in the regime of  subcritical $\Rey$. Berger et al. \cite{berger_coherent_1990} investigated the near-wake ($x/D \leq 9$) structure of a sphere and a disk using cross-spectral analysis and smoke visualization. Three frequencies dominated the near wake of a disk: a low-frequency ($\Str = 0.05$) axisymmetric ($m=0$)  pumping of the recirculation bubble, the vortex shedding frequency ($\Str = 0.135$) dominated by the helical $m=1$ mode, and a high frequency of ($\Str = 1.62$) related to the instability of the separated shear layer. The wake of a sphere at subcritical $\Rey$ was found to be similar to that of a disk. Cannon et al. \cite{cannon1993observations} found that these large-scale helical $m=1$ structures persisted even further downstream until $x/D = 29$. Later, Johansson et al. \cite{johansson_proper_2002} analyzed the wake of disk at $\Rey = 26,400$ using hot-wire measurements of $u'_x$ and proper orthogonal decomposition (POD) \cite{lumley_1967, lumley_1970}. Two distinct peaks were present in their POD spectra: (i) $m=1$, $\Str = 0.126$ associated with vortex shedding from the disk, and (ii) $m=2$, $\Str \approx 0$. 
 They found  that the $m=2$ mode eventually dominated the energy content of the wake by $x/D = 50$, where $D$ is the diameter of disk. This study was later extended  by Johansson and George \cite{johansson_far_2006-1} who performed measurements until $x/D = 150$. They found the dominance of $m=2$ appeared at  $x/D \approx 30$ beyond which the turbulence statistics also started exhibiting self-similar behavior.
 
Different from  experiments, attempts to study the evolution of coherent structures in turbulent wakes using numerical simulations have mainly relied on flow visualizations. Constantinescu and Squires \cite{constantinescu_les_2003} used the vortex identification method proposed by Jeong and Hussain \cite{jeong_identification_1995} to visualize the coherent structures in the wake of sphere at $\Rey = 10,000$. They observed that the main coherent structure shed patches of vorticity which then rotated irregularly while being convected downstream. Yun et al. \cite{yun_vortical_2006} used the method of Jeong and Hussain \cite{jeong_identification_1995} in conjunction with particle tracking to study the vortex structure of sphere wake at $\Rey = 3700$ and $10,000$. At lower $\Rey = 3700$, the separated shear layer formed a cylindrical vortex sheet becoming unstable at $x/D \approx 2$. At $\Rey= 10,000$, the separated shear layer became unstable immediately behind the body and  formed vortex rings. Using  particle tracking, they showed that the helical structure of the wake was not due to the rotation of vortical structures in the azimuthal direction but was due to a helical mode that translated downstream without rotation. This finding was further confirmed in later studies of Rodriguez et al. \cite{rodriguez_direct_2011} at   $\Rey = 3700$ and Chongsiripinyo and Sarkar \cite{chongsiripinyo_effect_nodate} at  $\Rey = 10,000$. 

In the classical formulation of POD proposed by Lumley \cite{lumley_1967, lumley_1970}, the homogeneous directions are first separated using the Fourier transform and then the cross-spectral tensor is decomposed in the non-homogeneous directions to give POD eigenvalues and eigenmodes. Thus, for  statistically stationary flows, each POD mode is characterized by a single frequency. Since its introduction to the fluid dynamics community, this original formulation of POD has been extensively used by experimentalists to educe  coherent structures in different types of turbulent flows. In the context of free shear flows, Leib et al. \cite{leib1984coherent}, Glauser et al. \cite{glauser1987coherent}, and Glauser and George \cite{glauser1987orthogonal}  applied the classical POD to the near-field measurements of a turbulent axisymmetric jet.  Thereafter, there have been several experimental studies employing the classical form of POD to study  coherent structures in a variety of flow configurations: (i) turbulent jets \cite{bonnet_stochastic_1994,arndt_proper_1997, gordeyev_coherent_2000,citriniti_reconstruction_2000, gordeyev_coherent_2002, iqbal_coherent_2007, davoust_dynamics_2012}, (ii) mixing layers \cite{bonnet_stochastic_1994,delville_examination_1999, ukeiley_examination_2001}, and (iv) wakes \cite{johansson_proper_2002, johansson_far_2006-1, tutkun_three-component_2008}.

There have been some POD studies of simulation data from various flows but POD analysis of turbulent wake simulations is lacking. Simulation-based POD  been dominated by its `snapshot'-type variant introduced by Sirovich \cite{sirovich1987turbulence}. In  snapshot POD \cite{sirovich1987turbulence}, the spatial correlation tensor is decomposed and the modes possess spatial coherence while evolving randomly in time. As a result, the snapshot POD modes are generally not coherent in time. On the other hand, application of classical POD for numerical simulations require long time integration making its application challenging for large-scale computations. Recently Towne et al. \cite{towne_spectral_2018} revisited a form of POD that leverages the temporal symmetry of statistically stationary flows termed spectral POD (SPOD). SPOD has been extensively used to study the coherent structures in compressible jets and their link to noise generation \cite{schmidt_wavepackets_2017, schmidt_spectral_2018, nogueira_large-scale_2019,lesshafft_resolvent-based_2019}.

The main objective of the present study is to improve upon our previous understanding of the coherent structures in the wake of a disk.  This is achieved by conducting an extensive SPOD analysis of  data from flow past a  disk simulated at $\Rey = 50,000$ by Chongsiripinyo and Sarkar~\cite{chongsiripinyo_decay_2020}, specifically their case of the wake in a homogeneous, unstratified fluid. We improve on previous experimental studies by including all three velocity components instead of solely $u_x$, by employing the  high spatial resolution and coverage  possible with simulation data,  by analyzing the flow field at  several downstream locations from near the body to the far wake, and by considering a higher $\Rey$. The ability of SPOD to separate temporal and spatial scales makes it a desirable candidate to study the coherent structures in turbulent flows. It is hoped that the qualitative findings of this study will be applicable to the wakes of other bluff bodies too, e.g., a sphere. 

While the previous experimental studies of the wake of a disk using POD have developed our understanding of the role of the $m=1$ vortex shedding mode and the  dominance of the $m=2$ double helix mode away from the body, a more complete analysis of the eigenspectrum and eigenmodes of the  azimuthal modes including their relative importance is missing. We bridge this gap by analyzing the SPOD eigenmodes and eigenspectra of different azimuthal modes in detail. Specifically, we consider the following questions. How is the energy distributed among SPOD modes of different azimuthal wave numbers ($m$) and frequencies ($\Str$), both in the near as well as the far wake? Does the turbulent wake of a disk at a higher $\Rey$, as in the present case, show the dominance of the $m=2$ azimuthal mode akin to the experimental investigations of the past? Do the SPOD eigenvalues and eigenmodes of the different dominant modes exhibit self-similarity indicating their connection to the local turbulence structure instead of being global modes?

We also explore the reconstruction of the turbulent kinetic energy (TKE) and $\langle u'_{x}u'_{r} \rangle$ to further clarify the role of modal decomposition. In particular, we address the following questions. What is the distribution of TKE and $\langle u'_{x}u'_{r} \rangle$ in the leading SPOD modes of the dominant azimuthal wave numbers and frequencies? How does the reconstruction of both TKE and $\langle u'_{x}u'_{r} \rangle$ change when we systematically change the reconstruction parameters by varying $m$, $\Str$ and the number of SPOD modes?  

The remainder of the paper is organized as follows. In section \ref{numerical_details} and \ref{spod_overview}, we present the numerical methodology and a brief description of SPOD, respectively. Some visualizations of the flow follow in section \ref{visualization}. Section \ref{uxur_k_real_data} is a presentation of  some single-point statistics obtained from ensemble-averaging (averaging in time and in the azimuthal direction) of the numerical data. Section \ref{eigenvalues_section} and \ref{spod_eigenmodes} is a description of SPOD eigenvalues and eigenmodes  at locations $x/D \geq 20$. The sensitivity of the reconstruction of TKE and $\langle u'_{x}u'_{r} \rangle$ to  the selection of SPOD modes is discussed in  section \ref{tke_uxur_recon}. Finally, we report the SPOD analysis at a 
few locations near  the body in Section \ref{spod_nearbody} and present conclusions in section \ref{conclusions}. 

\section{Governing equations and numerical scheme}\label{numerical_details}

The flow past a disk in a homogeneous fluid was simulated at $\Rey = 50,000$.  As reported by \cite{chongsiripinyo_decay_2020},  a  large eddy simulation (LES) approach was adopted and the simulation was conducted with high resolution.  The non-dimensional filtered Navier-Stokes equations governing the flow are as follows:

continuity:
\begin{equation} 
\frac{\partial u_{i}}{\partial x_{i}} = 0,
\label{conservation_eqn}
\end{equation}

momentum:

\begin{equation} 
\frac{\partial u_{i}}{\partial t} + \frac{\partial (u_{i}u_{j})}{\partial x_{j}} = -\frac{\partial p}{\partial x_{i}} + \frac{1}{\Rey}\frac{\partial}{\partial x_{j}}\Big[\Big(1 + \frac{\nu_{s}}{\nu}\Big)\frac{\partial u_{i}}{\partial x_{j}}\Big], 
\label{momentum_eqn}
\end{equation}
where $u_{i}$ corresponding to $i = 1, 2$, and $3$ refers to filtered fluid velocities in streamwise ($x_{1}$), lateral ($x_{2}$), and vertical ($x_{3}$) directions, respectively. In Eq. (\ref{momentum_eqn}), $\nu_{s}$ and $\nu$ refer to the kinematic subgrid viscosity obtained from the LES formulation and the kinematic viscosity of the fluid, respectively. The governing equations are non-dimensionalized using the following parameters: free-stream velocity ($U_{\infty}$) for velocity, diameter of disk ($D$) for spatial locations ($x_{i}$), dynamic pressure ($\rho_{o}U_{\infty}^{2}$) for pressure ($p$), and advection time ($D/U_{\infty}$) for time ($t$). The Reynolds number is denoted by $\Rey = U_{\infty}D/\nu$.

The filtered Navier-Stokes equations given by Eq. (\ref{conservation_eqn}) and Eq. (\ref{momentum_eqn}) are solved in a cylindrical coordinate system for the streamwise axial velocity ($u_{x}$), radial velocity ($u_{r}$), azimuthal velocity ($u_{\theta}$) and pressure ($p$). The field variables are functions of streamwise location ($x$), radial distance from the axis ($r$), and azimuthal location ($\theta$). The disk is centered at $(x_{1}, x_{2},x_{3})=(0,0,0)$  in the computational domain, and is represented by the immersed boundary method of Balaras \cite{balaras_modeling_2004}; Yang and Balaras \cite{yang_embedded-boundary_2006}. Spatial derivatives are computed using second-order accurate finite central differences. The temporal marching is performed using the fractional step method which combines the low-storage Runge-Kutta-Wray (RKW3) scheme with the second-order Crank-Nicolson scheme. Taking the divergence of velocity in the predictor step, a pressure Poisson equation is formed which, after taking account of periodicity in the azimuthal direction,  transforms to a linear system of equations for Fourier pressure modes. The linear system involves a pentadiagonal matrix which  is inverted using a direct solver \cite{rossi_parallel_1999}. The kinematic subgrid viscosity ($\nu_{s}$) is obtained using  the dynamic  eddy viscosity model of  Germano et al. \cite{germano_dynamic_1991}. At the inlet boundary, a uniform stream of velocity ($U_{\infty}$) is imposed while an Orlanski-type convective boundary condition is used for the outflow \cite{orlanski_simple_1976}. Neumann boundary condition is imposed at the radial boundary of the domain for all three velocity components.

The computational domain for the present simulation extends until  $x/D = 125$ in  the streamwise and $r/D = 15$ in the radial direction. The  number of grids points used to discretize the domain is as follows: $N_{r} = 364$ in the radial direction, $N_{\theta} = 256$ in the azimuthal direction, and $N_{x} = 4608$ in the axial direction. This  choice results in approximately 430 million elements. The conducted LES has high resolution. At $x/D = 10$, $\Delta x/ \eta$ is smaller than 10, decreasing to below 6 by $x/D = 125$. The resolution in the other directions is similarly good. Chongsiripinyo and Sarkar \cite{chongsiripinyo_decay_2020} can be referred for more details regarding the numerics.

\section{Description of Spectral Proper Orthogonal Decomposition (SPOD)} \label{spod_overview}

\subsection{Overview of POD for statistically stationary flows}
Let us consider a zero-mean stochastic process $\mathbf{u}(\textbf{x},t)$ in a finite spatial domain $\Omega$. In the context of turbulent flows, $\mathbf{u}(\textbf{x},t)$ can be considered as the fluctuating component of the full velocity field. POD proposed by Lumley \cite{lumley_1967, lumley_1970} aims at obtaining deterministic functions $\mathbf{\Psi}(\mathbf{x},t)$ 
on which $\mathbf{u}(\textbf{x},t)$ has the  maximum ensemble-averaged projection. Analytically, this maximization is expressed as, 
\begin{equation}
\max_{\boldsymbol{\Psi}}  \frac{\langle \{\mathbf{u}(\mathbf{x},t), \boldsymbol{\Psi}(\mathbf{x},t)\}\rangle}{||\mathbf{\Psi}(\mathbf{x},t)||^{2}},
\label{eq1}
\end{equation}
where $\langle . \rangle$ represents the ensemble average and the inner product $\{\mathbf{u}(\mathbf{x},t), \mathbf{v}(\mathbf{x},t)\}$ is defined as
\begin{equation}
\{\mathbf{u}(\mathbf{x},t), \mathbf{v}(\mathbf{x},t)\} = \int_{-\infty}^{\infty}\int_{\Omega} \mathbf{v}^{*}(\mathbf{x},t)\mathbf{W}(\mathbf{x})\mathbf{u}(\mathbf{x},t)\diff\mathbf{x}\diff t.
\label{eq2}
\end{equation}
Here, $\mathbf{W}(\mathbf{x})$ is a positive-definite Hermitian matrix and the asterisk denotes the complex conjugate of the vector field. Using the calculus of variation \cite{holmes2012turbulence}, the minimization of the expression in Eq. (\ref{eq1}) reduces to a Fredholm-type integral eigenvalue equation given by
\begin{equation} 
\int_{-\infty}^{\infty}\int_{\Omega}R_{ij}(\mathbf{x},\mathbf{x}',t,t')\mathbf{W}(\mathbf{x}')\Psi^{(n)}_{j}(\mathbf{x}',t')\diff\mathbf{x}'\diff t' = \lambda^{(n)} \Psi^{(n)}_{i}(\mathbf{x},t),
\label{eq3}
\end{equation}
where $\lambda^{(n)}$ and $\Psi^{(n)}_{i}(\mathbf{x},t)$ are the $n^{th}$ eigenvalue and the component of the corresponding eigenmode in the $i^{th}$ direction, respectively. In Eq. (\ref{eq3}), $R_{ij}(\mathbf{x}, \mathbf{x'}, t, t') = \langle u_{i}(\mathbf{x},t)u_{j}^{*}(\mathbf{x}',t')\rangle$ corresponds to the space-time cross-correlation tensor.  

Since time ($t$) is a homogeneous direction in statistically stationary flows, $R_{ij}(\mathbf{x}, \mathbf{x'}, t, t')$  for such flows can be written as, 
\begin{equation}
R_{ij}(\mathbf{x}, \mathbf{x'}, t, t') = R_{ij}(\mathbf{x}, \mathbf{x'}, \tau) = \int_{-\infty}^{\infty} S_{ij}(\mathbf{x}, \mathbf{x'}, f)e^{-i2\pi f\tau}\diff f,
\label{correlation_tensor_fft}
\end{equation}
where $\tau = t - t'$ and $S_{ij}(\mathbf{x}, \mathbf{x'}, f)$ is the Fourier transform of $R_{ij}(\mathbf{x}, \mathbf{x'}, \tau)$. Using Eq. (\ref{correlation_tensor_fft}), the eigenvalue problem given by Eq. (\ref{eq3}) can be recast as the following equivalent problem \cite{towne_spectral_2018},
\begin{equation}
\int_{\Omega}S_{ij}(\mathbf{x},\mathbf{x}',f)\mathbf{W}(\mathbf{x}')\Phi^{(n)}_{j}(\mathbf{x}',f)\diff\mathbf{x}' = \lambda^{(n)}(f) \Phi^{(n)}_{i}(\mathbf{x},f),
\end{equation}
which can be solved at each frequency $f$. The modified eigenmodes  are then given by $\Phi^{(n)}_{i}(\mathbf{x},f) = \Psi^{(n)}_{i}(\mathbf{x},t)e^{-i2\pi f t}$. By virtue of the Hilbert-Schmidt theorem, the eigenvalues are sorted such that $\lambda^{(1)}(f)  \geq \lambda^{(2)}(f) \geq . \ . \ . \geq \lambda^{(n)}(f) $ where $\lambda^{(n)}(f)$ represents the energy content of the $n^{th}$ mode at the frequency $f$. The eigenmodes are orthonormal to each other, i.e.,
\begin{equation}
\int_{\Omega} \mathbf{\Phi}^{*(n)}(\mathbf{x},f)\mathbf{W}(\mathbf{x})\mathbf{\Phi}^{(m)}(\mathbf{x},f)\diff \mathbf{x} = \delta_{mn}, 
\label{eq4}
\end{equation}
where $\delta_{mn}$ is the Dirac-delta function. These eigenmodes also provide a complete basis for the Fourier realization of the turbulent velocity field $\mathbf{u}(\mathbf{x},t)$ at frequency $f$, i.e.,
\begin{equation}
\hat{\mathbf{u}}(\mathbf{x},f) = \sum\limits_{n=1}^{\infty}a^{(n)}(f)\mathbf{\Phi}^{(n)}(\mathbf{x},f),
\end{equation}
where $a^{(n)}(f) = \{\hat{\mathbf{u}}(\mathbf{x},f), \mathbf{\Phi}^{(n)}(\mathbf{x},f)\}$ is the inner product of the Fourier transform of $\mathbf{u}(\mathbf{x},t)$ and the $n^{th}$ eigenmode at frequency $f$.

\subsection{Numerical implementation of SPOD}\label{num_spod}

In the present work, SPOD is applied to two-dimensional (2D) cross-stream slices of the three-dimensional (3D) velocity field sampled at different streamwise locations from the numerical simulation. Downstream locations, ranging from $x/D = 5$ to $100$, are sampled at  a spacing of approximately $5D$. Two additional locations at $x/D = 110$ and $120$ are also sampled. Besides these locations, SPOD is also performed at $x/D = 0.1, 1, 2 $, and $5$ to analyze the modal distribution of fluctuation energy  near  the disk. 

The turbulent wake behind a disk is homogeneous-periodic in the azimuthal direction. It can be shown that the SPOD eigenfunctions in the azimuthal direction (or any other homogeneous direction for that matter) are harmonic functions \cite{lumley_1970, towne_spectral_2018}. Owing to the statistically stationary nature of the wake, the azimuthally decomposed velocity field can be further decomposed into the temporal Fourier modes such that
\begin{equation}
\mathbf{u}(x;r,\theta,t) = \sum_{m}\mathbf{\tilde{u}}_{m}(x;r,t)e^{im\theta} = \sum_{f}\sum_{m}\mathbf{\hat{u}}_{mf}(x;r)e^{im\theta}e^{i2\pi ft},
\end{equation}
where $\mathbf{\hat{u}}_{mf}$ is the double Fourier decomposed velocity field for a given $(m,f)$ pair and $\mathbf{\tilde{u}}_{m}$ is the azimuthally decomposed instantaneous snapshot at a time instant $t$ . 

For the numerical implementation of the SPOD, the velocity field is first decomposed in  the azimuthal direction and the data for each azimuthal mode is  collected into a snapshot matrix $\mathbf{U}_{m}$ as
\begin{equation}
\mathbf{U}_{m} = [\mathbf{\tilde{u}}_{m}^{(1)} \mathbf{\tilde{u}}_{m}^{(2)} \cdots \mathbf{\tilde{u}}_{m}^{(N)}],
\end{equation}
where $N$ is the total number of time snapshots used for the SPOD. Subsequently, $\mathbf{U}_{m}$ is divided into $N_{blk}$ overlapping blocks, each  containing $N_{freq}$ entries, as follows:
\begin{equation}
\mathbf{U}_{m}^{(l)} = [\mathbf{\tilde{u}}_{m}^{(l)(1)} \mathbf{\tilde{u}}_{m}^{(l)(2)} \cdots \mathbf{\tilde{u}}_{m}^{(l)(N_{freq})}],
\end{equation}
where $\mathbf{U}_{m}^{(l)}$ is the $l^{th}$ block consisting of $N_{freq}$ time snapshots. Each block is then Fourier transformed in the temporal direction and all realizations at a given frequency $f$ are collected into a matrix $\mathbf{\hat{U}}_{mf}$ as
\begin{equation}
\mathbf{\hat{U}}_{mf} = [\mathbf{\hat{u}}_{mf}^{(1)} \mathbf{\hat{u}}_{mf}^{(2)} \cdots \mathbf{\hat{u}}_{mf}^{(N_{blk})}].
\end{equation}

At this stage, the $(r,\theta, t)$ simulation  data  at each of  the chosen streamwise planes has been represented as a collection of $N_{blk}$ independent realizations of the $(r, m,f)$ dependence of the three velocity components. From this form of the data, eigenvectors and eigenvalues are obtained by the eigenvalue decomposition of the weighted cross-spectral density matrix:
\begin{equation}	
\mathbf{\hat{U}}_{mf}^{*}\mathbf{W}\mathbf{\hat{U}}_{mf} \mathbf{\Gamma}_{mf} = \mathbf{\Lambda}_{mf}\mathbf{\Gamma}_{mf}. 
\end{equation}

Here, $\mathbf{W}$ is a $3N_{r} \times 3N_{r}$ diagonal matrix which contains the quadrature weights of radial grid points for all three velocity components, accounting for the numerical area-integration of TKE on the discrete grid. This ensures that the obtained SPOD modes optimally capture the area-integrated TKE at any $x/D$ location. The SPOD modes for a given $(m,f)$ can then be obtained from the eigenvectors $\mathbf{\Gamma}_{mf}$ as $\mathbf{\Phi}_{mf} = \mathbf{\hat{U}}_{mf}\mathbf{\Gamma}_{mf}\mathbf{\Lambda}_{mf}^{-1/2}$. The obtained eigenmodes are orthogonal and the eigenvalues are ordered with respect to their contribution to area-integrated fluctuation kinetic energy as described in the previous section. 

For the present analysis, $N=7200$ snapshots are used for the analysis. Consecutive snapshots are separated by non-dimensional time $\Delta t D/U_{\infty} \approx 0.07$. $N_{freq}$ (size of each block) and $N_{blk}$ (overlap between two consecutive blocks) are set as $512$ and $256$ respectively, resulting in total of $N_{blk} = 27$ SPOD modes for each pair of $(m,f)$. Thus, in the present application of SPOD, $\mathbf{\hat{U}}_{mf}$ is a matrix of dimension $3N_{r} \times N_{blk}$. It is worth noting that  one block, consisting of $N_{freq} = 512$ snapshots, spans a time window $T_{block} = 36.91D/U_{\infty}$. The integral timescale  at $r/D = 0.5$, evaluated by integrating the auto-correlation function of streamwise fluctuation velocity ($u'_{x}$) from a zero value of  time lag ($\tau$)   to the first zero crossing \cite{katul_analysis_1995,oneill_autocorrelation_nodate}, varies from $\gamma = 0.6045D/U_{\infty}$ at $x/D =10$ to $\gamma = 1.6529D/U_{\infty}$ at $x/D = 120$. Thus, at $x/D = 10$, one block of 512 snapshots spans approximately  61 integral timescales which decreases to approximately 22 integral timescales by $x/D = 120$. Readers are referred to Towne et al. \cite{towne_spectral_2018} for more details regarding SPOD and its connection to different modal decomposition techniques (e.g., DMD, resolvent analysis, etc.) and to Schmidt and Colonius \cite{schmidt2020guide} for an introduction to the method.

\section{Visualizations} \label{visualization}

\begin{figure}
\centering
\includegraphics[width=\linewidth, keepaspectratio]{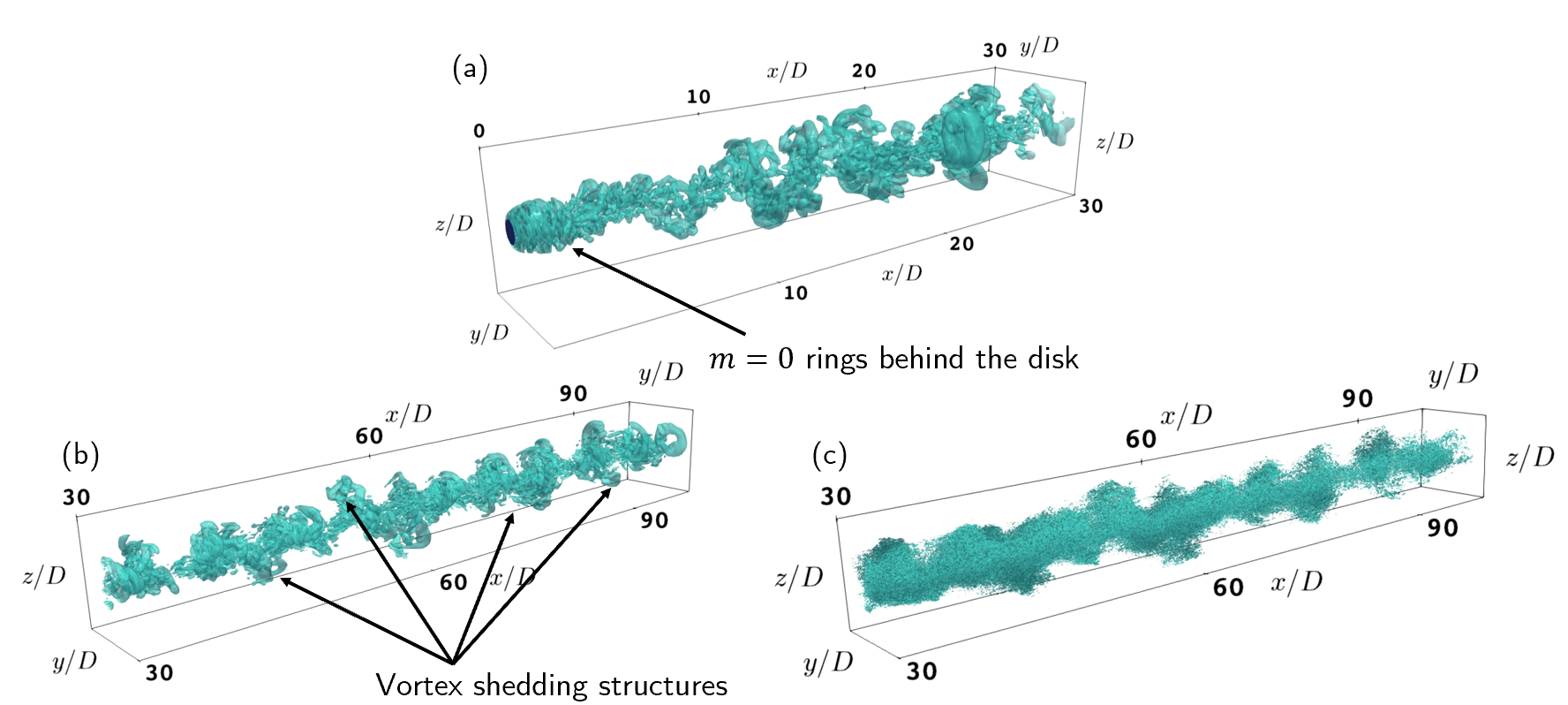}
\caption{Isosurfaces of $Q$-criterion. $Q=0.001$ of the  filtered velocity for (a) $0 < x/D < 30$ and for (b) $30 < x/D < 100$ at a given time instant; (c) $Q=0.05$ of the residual field for $30 < x/D < 100$.}
\label{fig:qcri_3d}
\end{figure}

Figure \ref{fig:qcri_3d} shows three-dimensional instantaneous visualizations of Q-criterion
\cite{hunt_ctr_1988}, which is used to identify vorticity-dominated regions in a flow field. $Q$ is the second invariant of the velocity gradient tensor, defined as
\begin{equation}
Q = \frac{1}{2}(|\mathbf{\Omega}^{2}| - |\mathbf{S}|^{2}),
\end{equation}
where
\begin{equation}
\Omega_{ij} = \frac{1}{2}\Big(\frac{\partial u_{i}}{\partial x_{j}} - \frac{\partial u_{j}}{\partial x_{i}}\Big), \quad
 S_{ij} = \frac{1}{2}\Big(\frac{\partial u_{i}}{\partial x_{j}} + \frac{\partial u_{j}}{\partial x_{i}}\Big)
\end{equation}
are the rotation tensor and strain-rate tensor respectively. Regions with $Q>0$ are dominated by vorticity signifying that the fluid motion is primarily rotational in those regions.  

At the high $\Rey$ of the present study, velocity gradients are found to be dominated by small-scale turbulent fluctuations. To focus on the large-scale coherent structures, the instantaneous velocity field is filtered using a Gaussian low-pass filter, an in-built SciPy function named \textit{gaussian\textunderscore filter}. In the inputs for the function \textit{gaussian\textunderscore filter}, the standard deviation ($\sigma$) of the Gaussian kernel was varied systematically from $\sigma = 2$ to $30$. Subsequently, based on  visual inspection, a Gaussian low-pass filter with $\sigma = 10$ was used for the present visualizations. Higher $\sigma$ values led to the smearing of large-scale coherent structures while visualizations with lower $\sigma$ still had significant imprints of the small-scale turbulence obscuring the large-scale coherent structures. The width of the Gaussian kernel is set such that the $Q$ of the filtered velocity field elucidates coherent structures without much distortion and, at the same time, is not completely dominated by the small-scale fluctuations. 

\begin{figure}
\centering
\includegraphics[width=\linewidth, keepaspectratio]{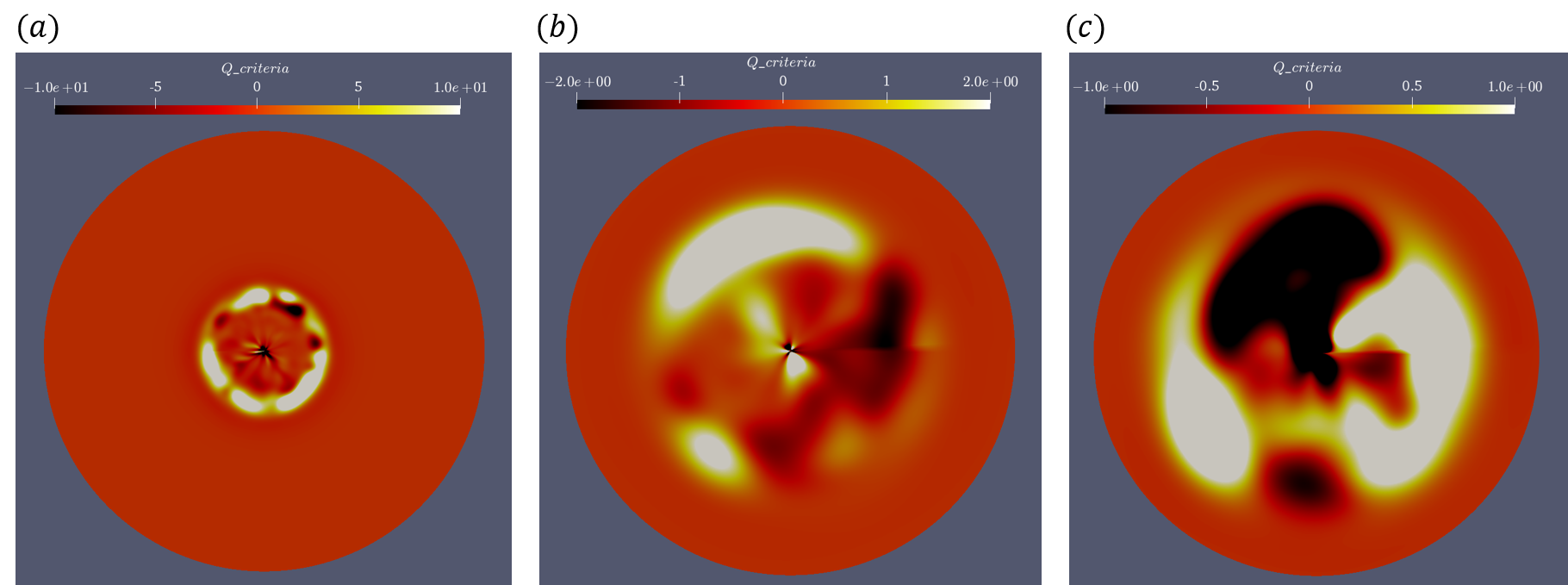}
\caption{Two-dimensional contours of $Q$-criterion of the filtered velocity fields at the same time instant as in Fig. \ref{fig:qcri_3d} at various streamwise locations: (a) $x/D = 0.05$, (b) $x/D = 0.5$, and (c) $x/D = 1.05$.}
\label{fig:qcri_azimuthal}
\end{figure}

In Fig. \ref{fig:qcri_3d}(a), it can be observed that vortex rings are shed in the immediate downstream of the disk. These vortex rings represent the axisymmetric $m=0$ mode. As the flow evolves spatially, these rings become unstable and give way to a complex distribution of vorticity in the wake. Instantaneous two-dimensional contours of $Q$ in the vicinity of the disk (see Fig. \ref{fig:qcri_azimuthal}) shows  higher azimuthal modes with $m>0$ (see Fig. \ref{fig:qcri_azimuthal}(a)) that emerge close to the disk and distort the $m=0$ vortex rings. At $x/D = 0.5$ and $1.05$, the presence of $m=1$ and $m=2$ modes can be seen in Fig. \ref{fig:qcri_azimuthal}(b) and \ref{fig:qcri_azimuthal}(c), respectively. 

Despite the entangled arrangement of vortices, a helical orientation of coherent structures in the wake can be discerned from the 3D visualizations of Fig. \ref{fig:qcri_3d}. These coherent structures are the vortex shedding structures that originate near the disk from instabilities in $m=0$ and advect downstream. Visual inspection of Fig. \ref{fig:qcri_3d}(a) and (b) reveal the following points. First, the vortex shedding structures are separated approximately by $\lambda_{VS}/D = 1/\Str$ where $\Str=0.135$ (identified  formally by the modal decomposition, as will be seen) is the vortex shedding frequency of the disk wake at hand. Second, these structures meander away from the wake centerline as the flow evolves downstream. For completeness, $Q$-criterion of the residual velocity field, obtained by subtracting the filtered velocity from the original velocity, is presented in Fig. \ref{fig:qcri_3d}(c) for $30 < x/D < 100$. The residual field also shows a helical-like orientation similar to the filtered field observed in Fig. \ref{fig:qcri_3d}(b). It is worth noting that the above-mentioned procedure of obtaining the residual field does not ensure the absence of an imprint of the  large-scale features on  the residual field. It is also possible that this imprint can be physical (rather than the imperfection of scale separation by a physical-space filter)  in the sense that some of the fine-scale turbulence is ``slaved" to the coherent structures. 

\section{Evolution of  turbulence statistics in the wake} \label{uxur_k_real_data}

\begin{figure}
\centering
\label{fig:turb_to_ud}
\includegraphics[width=0.7\linewidth]{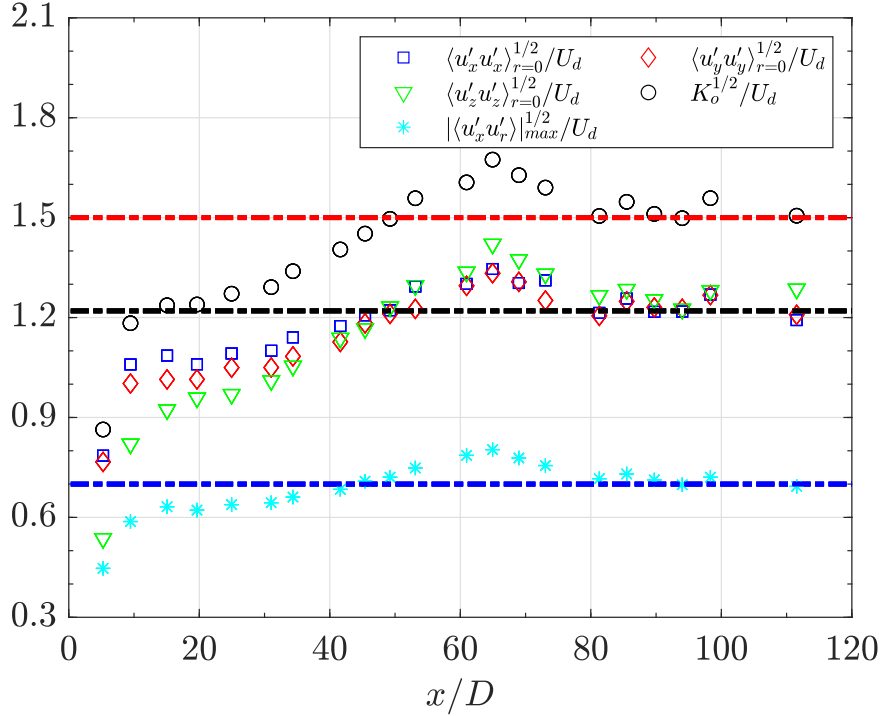}
\caption{Ratio of centerline turbulent intensities and $|\langle u'_{x}u'_{r} \rangle|^{1/2}_{max}$ to $U_{d}$ as a function of $x/D$.}
\label{fig:turb_quantities_centerline}
\end{figure}

Figure \ref{fig:turb_quantities_centerline} shows the streamwise evolution of 
 the centerline r.m.s. velocity fluctuations and the maximum value of $\langle -u'_{x}u'_{r} \rangle^{1/2}(r)$, each normalized by  the centerline defect velocity ($U_{d}$). The normalized turbulent velocity scale ($K_{o}^{1/2}$),  derived from the centerline TKE ($K_{o} = \langle u'_{i}u'_{i} \rangle_{r=0} / 2 $), is also shown. Chongsiripinyo and Sarkar \cite{chongsiripinyo_decay_2020} found, using the same simulation,  that the mean velocity scale ($U_{d}$) and the turbulence velocity scale ($K_{o}^{1/2}$) did not follow the same decay rates for $10 < x/D < 65$;  $U_{d}$ was $ \propto x^{-0.9}$ while  $K_{o}^{1/2} $ was $\propto x^{-0.7}$. After $x/D \approx 65$, the decay rates of both $U_{d}$ and $K_{o}^{1/2}$ became similar and close to the classical decay exponent of $-2/3$ for the axisymmetric turbulent wake. The consequences of this difference in the initial decay rates can be observed in Fig. \ref{fig:turb_quantities_centerline} where the ratio of each r.m.s. velocity fluctuation to $U_d$ keeps increasing until  $x/D \approx 65$. Beyond $x/D = 65$, the ratios drop down and asymptote to  approximately $1.2$ for the individual r.m.s. fluctuations and 1.5 for  $K_{o}^{1/2}$. It can also be seen that the near-wake turbulence ($x/D < 40$) is more anisotropic with the streamwise component dominating over the other two. It is beyond $x/D =40$ that the r.m.s. velocity fluctuations become more or less isotropic. Figure \ref{fig:turb_quantities_centerline} also shows the ratio of square-root of the maximum value of $\langle -u'_{x}u'_{r} \rangle(r)$ to $U_{d}$. This ratio also increases for $x/D < 65$, albeit slowly, compared to the ratios of  the r.m.s. fluctuations. After $x/D \approx 80$, the value of $\langle -u'_{x}u'_{r} \rangle^{1/2}_{\rm max}/U_d$ asymptotes to  $\approx 0.7$. Compared to the previous experimental studies of flow past a disk (\cite{johansson_equilibrium_2003}, Table 5.3 in \cite{pope2002}), the ratios of turbulent intensities to the defect velocity are slightly higher in the present case (an asymptotic value of $1.2$ instead of approximately $0.9-1.1$ found in the previous studies), which may be due to the relatively high $\Rey$ of the current study.
 
\begin{figure}
\centering
\includegraphics[width=\linewidth,keepaspectratio]{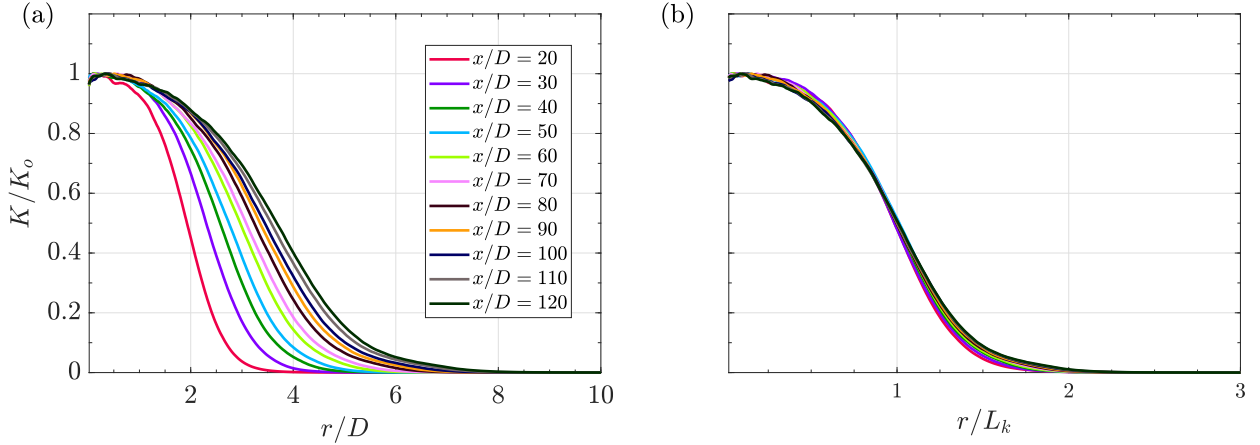}
\caption{TKE profiles scaled with $K_{o}$ at different streamwise locations ($20 < x/D < 120$) with radial direction scaled by: (a) $D$ and (b) $L_{k}$.}
\label{fig:tke_profiles}
\end{figure}

Figure \ref{fig:tke_profiles} shows radial profiles of TKE at different downstream locations spanning $20 < x/D < 100$. Normalization of  $r$ by the disk diameter ($D$) in part (a) is compared with normalization by the local wake width ($L_{k}$)  in part (b). Here,
$L_{k}$ is the half-width of the TKE profile defined by $K(x;r=L_{k}) = \frac{1}{2}K_{o}(x)$.
The downstream growth of wake thickness is seen in Fig. \ref{fig:tke_profiles}(a) where the radial spread of $K/K_{o}$ monotonically increases with increasing $x/D$. When the radial direction is scaled by $L_{k}$, these profiles collapse onto a single profile implying self-similar evolution of TKE beyond $x/D  \approx 20$. It is worth noting that the TKE  becomes approximately zero by $r/L_{k} = 2$.

\begin{figure}
\centering
\includegraphics[width=\linewidth, keepaspectratio]{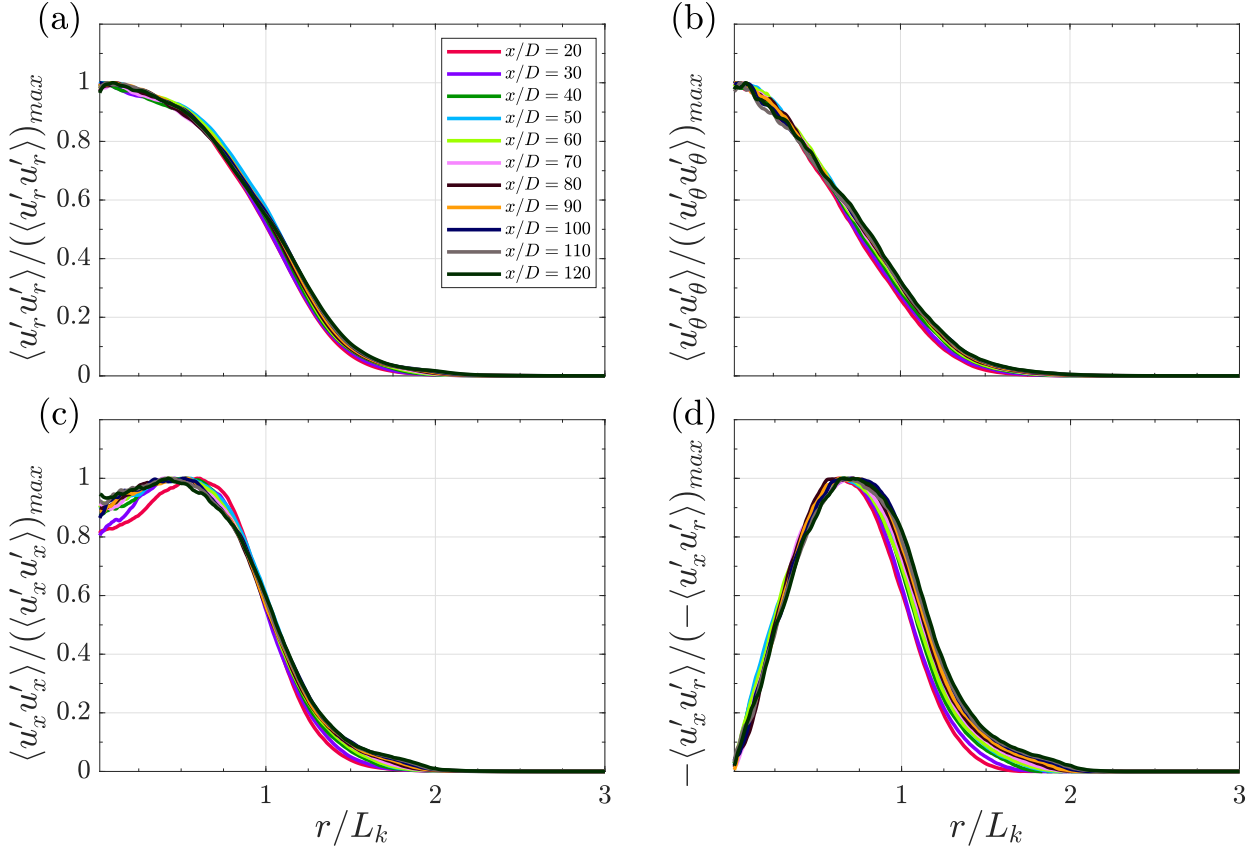}
\caption{Normal turbulent stresses (a, b, c) and $\langle u'_{x}u'_{r}\rangle$ (d) profiles at different streamwise locations ($20 < x/D < 120$) scaled by their maximum values at the respective locations. The radial direction is scaled by TKE-based wake width $L_{k}$.}
\label{fig:similarity_uxur_normal_stresses}
\end{figure}

In  Fig. \ref{fig:similarity_uxur_normal_stresses} we plot the scaled profiles of normal stresses along with $\langle u'_{x}u'_{r} \rangle$, the component of Reynolds stress tensor that appears in the simplified Reynolds-averaged streamwise momentum equation of the turbulent axisymmetric wake,

\begin{equation}
U_{\infty}\frac{\partial}{\partial x} (U-U_{\infty}) = -\frac{1}{r}\frac{\partial}{\partial r}(r\langle u'_{x}u'_{r} \rangle).
\label{rans_axisymmetric_wake}
\end{equation} 

All the turbulent stresses in Fig. \ref{fig:similarity_uxur_normal_stresses} have been scaled by their maximum values at the corresponding  $x/D$ locations, and the the radial direction has been scaled by the  TKE-based wake width $L_{k}$. All the three normal turbulent stresses plotted in Fig. \ref{fig:similarity_uxur_normal_stresses} collapse beyond $x/D \approx 40$. Until $x/D \approx 40$, the profile of the streamwise component shows the largest deviation among different locations. Besides $\langle u'_{x}u'_{x} \rangle$, which peaks between $r/L_{k} = 0.5-0.75$, the other two normal turbulent stresses peak near the centerline and decay with increasing $r$. All  three normal stresses approach zero by $r/L_{k} \approx 2$ as was also seen for the TKE profiles. 

The profiles of $\langle u'_{x}u'_{r} \rangle$ collapse well for $r/L_{k} < 0.75$ when scaled with $L_{k}$. Beyond the peak location of $\langle -u'_{x}u'_{r} \rangle$, which occurs near the peak of $\langle u'_{x}u'_{x} \rangle$, there is some spread in the normalized profiles. The radial extent of the scaled profiles increases with increasing $x/D$. At $x/D = 20$, the scaled Reynolds stress profile decays  to zero by $r/L_{k} \approx 1.6$. By $x/D = 100$, the radial extent of the scaled profiles has increased to $r/L_{k} \approx 2$.

\section{SPOD eigenvalues and eigenspectra} \label{eigenvalues_section}

\begin{figure}
\centering
\includegraphics[width=\linewidth]{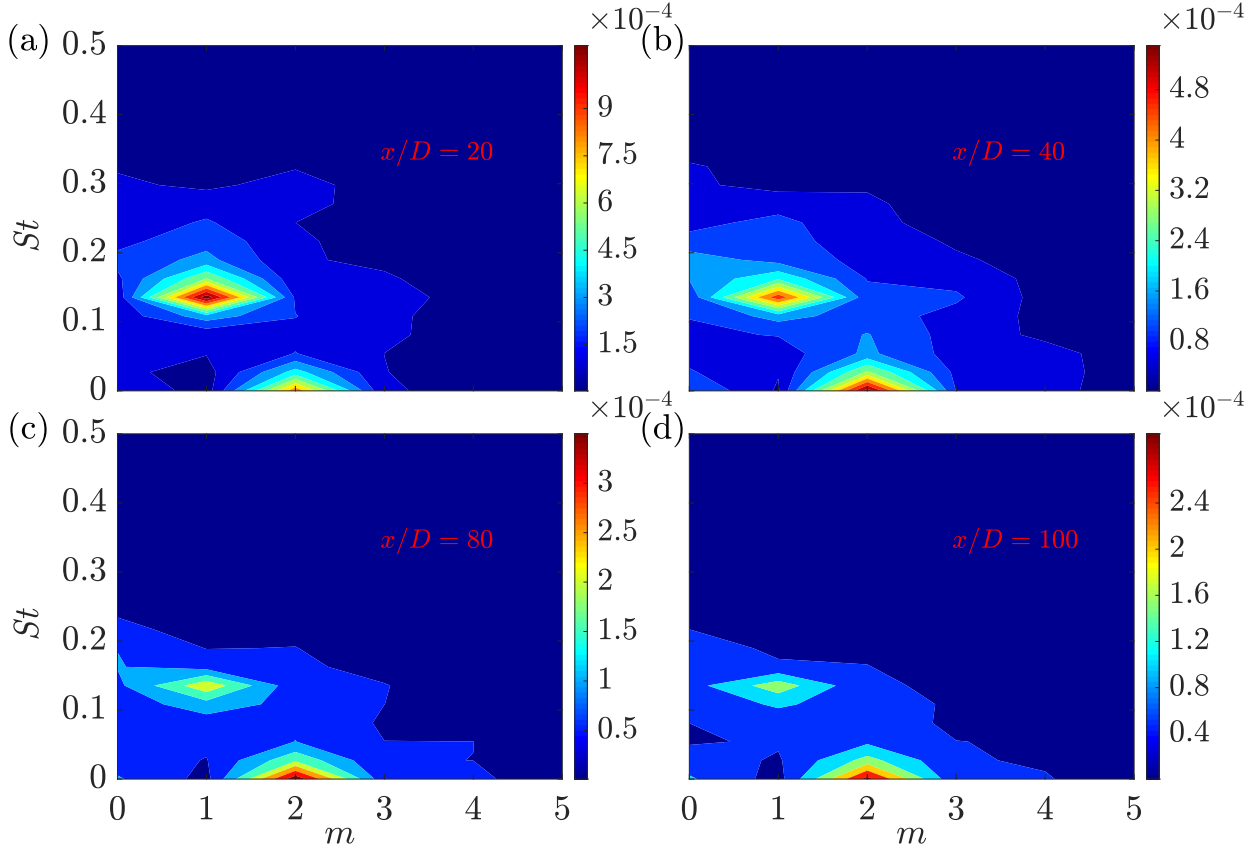}
\caption{SPOD contour maps showing energy contained in leading SPOD mode, $\lambda^{(1)}$, as a function of azimuthal wavenumber $m$ and frequency $\Str$ at different locations: (a) $x/D = 20$, (b) $x/D = 40$, (c) $x/D = 80$, and (d) $x/D = 100$. The colorbar limits are set according to the maximum values of $\lambda^{(1)}$ over all  ($m$, $\Str$) pairs at the respective $x/D$ locations.}
\label{fig:contour_maps_eigenvalues}
\end{figure}

Figure \ref{fig:contour_maps_eigenvalues} shows the distribution of energy in the leading SPOD (or the most energetic SPOD) eigenvalue $\lambda^{(1)}$ as a function of  azimuthal mode ($m$) and nondimensional frequency ($\Str$)  at four  downstream locations: $x/D = 20$, $40$, $80$, and $100$. At all locations, the energy in $\lambda^{(1)}$ among all $(m,\Str)$ pairs is predominantly contained  in modes that satisfy $m \leq 4$ and $St < 0.4$.

There are two distinct peaks in Fig. \ref{fig:contour_maps_eigenvalues}: (i) $m=1$, $\Str = 0.135$ (vortex shedding (VS) structure), and (ii) $m=2$, $\Str = 0$ (double helix (DH) structure). The former has long been known to be the vortex shedding structure in the turbulent wake of a disk \cite{fuchs_large-scale_1979,berger_coherent_1990, cannon1993observations, johansson_proper_2002, johansson_far_2006-1}. The existence of the latter in the high-$\Rey$ wake of disk was first reported by Fuchs et al. \cite{fuchs_large-scale_1979} and its importance was expanded upon later by Johansson and George \cite{johansson_far_2006-1}. The prominent peak at $\Str=0$ should be interpreted as a quasi-steady structure in the limit of $\Str \rightarrow 0$ \cite{nogueira_large-scale_2019}. The discrete nature of the Fourier transform and the limited temporal runtime ($T \approx 504D/U_{\infty}$, which is approximately four flow-through times) of these computationally intensive simulations make it difficult to resolve very small frequencies in the limit of $\Str \rightarrow 0$, leading to the energy of very low frequencies being captured in $\Str = 0$. In the rest of the paper, $\Str=0$ will be replaced with $\Str \rightarrow 0$ in the context of $m=2$ to avoid misinterpreting it as a temporally stationary mode.

Figure \ref{fig:contour_maps_eigenvalues} has two important implications. First, the peak associated with vortex shedding persists far downstream, being still present at $x/D = 100$. Second, the leading SPOD mode of the VS structure clearly dominates the near wake (at $x/D = 20$) and gradually declines in importance relative to the DH structure which eventually dominates the energy content in $\lambda^{(1)}$ by $x/D = 100$. This observation is consistent with the previous findings of Johansson and George \cite{johansson_far_2006-1} who found that the DH structure dominated in the wake beyond $x/D = 30$. 

\begin{figure}
\centering
\includegraphics[width=\linewidth]{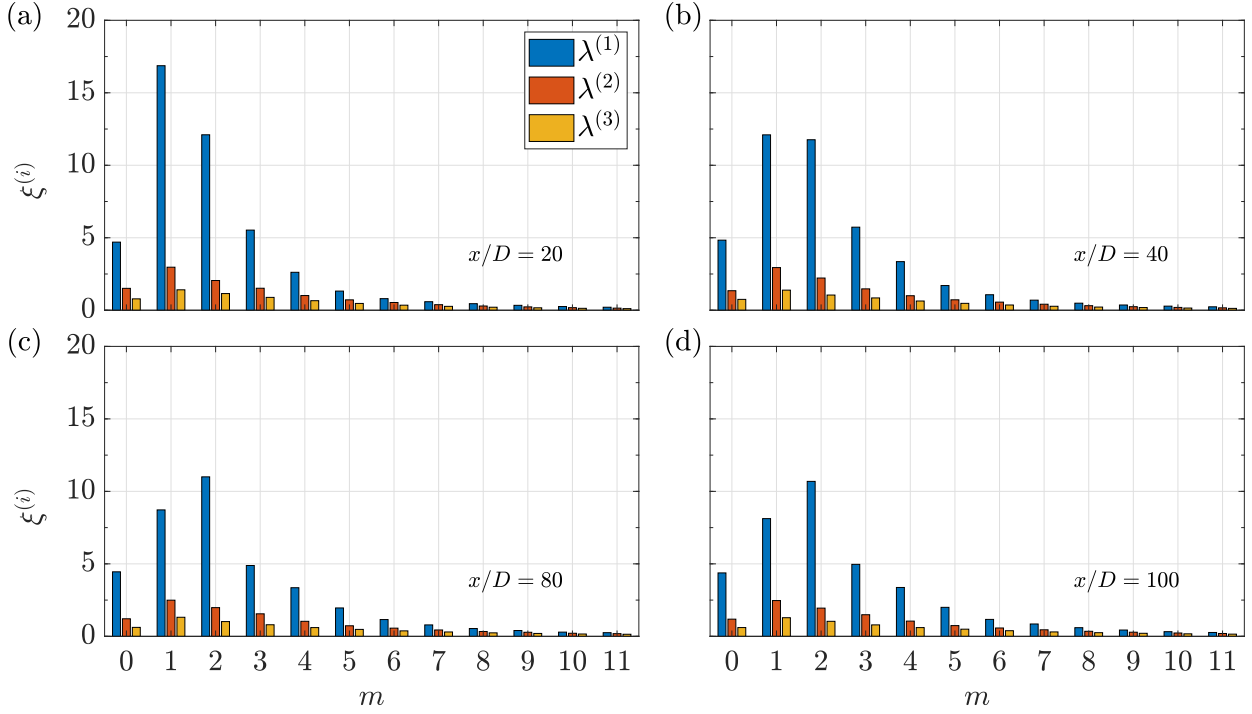}
\caption{Frequency-integrated eigenspectrum as a function of azimuthal mode number $m$ at different locations: (a) $x/D = 20$, (b) $x/D = 40$, (c) $x/D = 80$, and (d) $x/D = 100$. Three leading SPOD modes ($\lambda^{(1)}$, $\lambda^{(2)}$, and $\lambda^{(3)}$ ) at each $m$ are shown in terms of their percentage contributions to the area-integrated TKE.} 	
\label{fig:bar_plots_m}
\end{figure}

To further analyze the contribution of different azimuthal modes to the area-integrated TKE, the eigenspectrum of each $m$ has been summed over all  resolved frequencies and normalized with $E_{k}^{T} (x/D)$ to obtain the percentage contribution of each $m$ as follows:

\begin{equation}
\xi^{(i)}(m;x/D) = \frac{\displaystyle\sum_{\Str}\lambda^{(i)}(m,\Str;x/D)}{E_{k}^{T}(x/D)} \times 100,
\end{equation}
where the index $i$ corresponds to the $i^{th}$ SPOD mode and $E_{k}^{T} (x/D)$ is the  area-integrated TKE at that $x/D$ location.
The resulting frequency-integrated eigenspectrum has been plotted for four locations $x/D = 20, 40, 80$, and $100$ in Fig. \ref{fig:bar_plots_m}. From Fig. \ref{fig:bar_plots_m} it can be ascertained that a major contribution to TKE comes from the first five azimuthal modes in the near as well as far wake. Another observation is the overall low-rank behavior of azimuthal modes $m \le 4$ in the sense that there is a significant difference between the contributions of $\lambda^{(1)}$ and $\lambda^{(2)}$ for these $m$. At $x/D = 20$ in Fig. \ref{fig:bar_plots_m}(a), $m=1$ dominates the integrated eigenspectra followed by $m=2, 3$, and $0$ respectively. As $x/D$ increases, the relative contribution of $m=2$ starts increasing while that of $m=1$ starts declining. By $x/D = 40$ (Fig. \ref{fig:bar_plots_m}(b)), both $m=1$ and $m=2$ have similar contribution and eventually $m=2$ starts dominating the integrated eigenspectra as seen in Fig. \ref{fig:bar_plots_m}(c) and Fig. \ref{fig:bar_plots_m}(d). Beyond $m=2$, the energy content of $\lambda^{(i)}$ decreases monotonically with increasing $m$.

The results of Johansson and George \cite{johansson_far_2006-1} showed the eventual dominance of the $m=2$ mode beyond $x/D = 40$. In the present analysis, the $m=2$ mode emerges as the dominant mode at a farther downstream distance $x/D = 60$. It is worth noting that all three velocity components are included in the SPOD kernel as opposed to the  previous analysis  \cite{johansson_far_2006-1} which only included the streamwise velocity component.
Besides this difference, the present results show that the axisymmetric mode $m=0$ is always significantly less dominant than $m=1$ and is of comparable magnitude to the  $m=3$ mode. In the previous results \cite{johansson_far_2006-1}, the axisymmetric mode was of comparable magnitude to $m=1$ and was significantly more dominant than $m=3$ for all measurement stations at $30 \le x/D \le 150$ (see Fig. 7 in their paper). It is also worth noting that  $\Rey = 50,000$ is almost twice that of the previous study.

The findings of Fig. \ref{fig:contour_maps_eigenvalues} and \ref{fig:bar_plots_m} warrant a detailed investigation of the VS and DH mode.  In what follows, we investigate the $m=1$ and $m=2$ modes in more detail, particularly in the context of the VS and DH modes. We also present some results on the eigenvalues and eigenspectra of $m=0,3, $ and 4 modes since Fig. \ref{fig:bar_plots_m} shows that these modes, although not dominant, also make appreciable contributions  to the area-integrated TKE.

\subsection{Eigenspectra of $m=1$ and $m=2$ modes} 

\begin{figure}
\centering
\includegraphics[width=\linewidth]{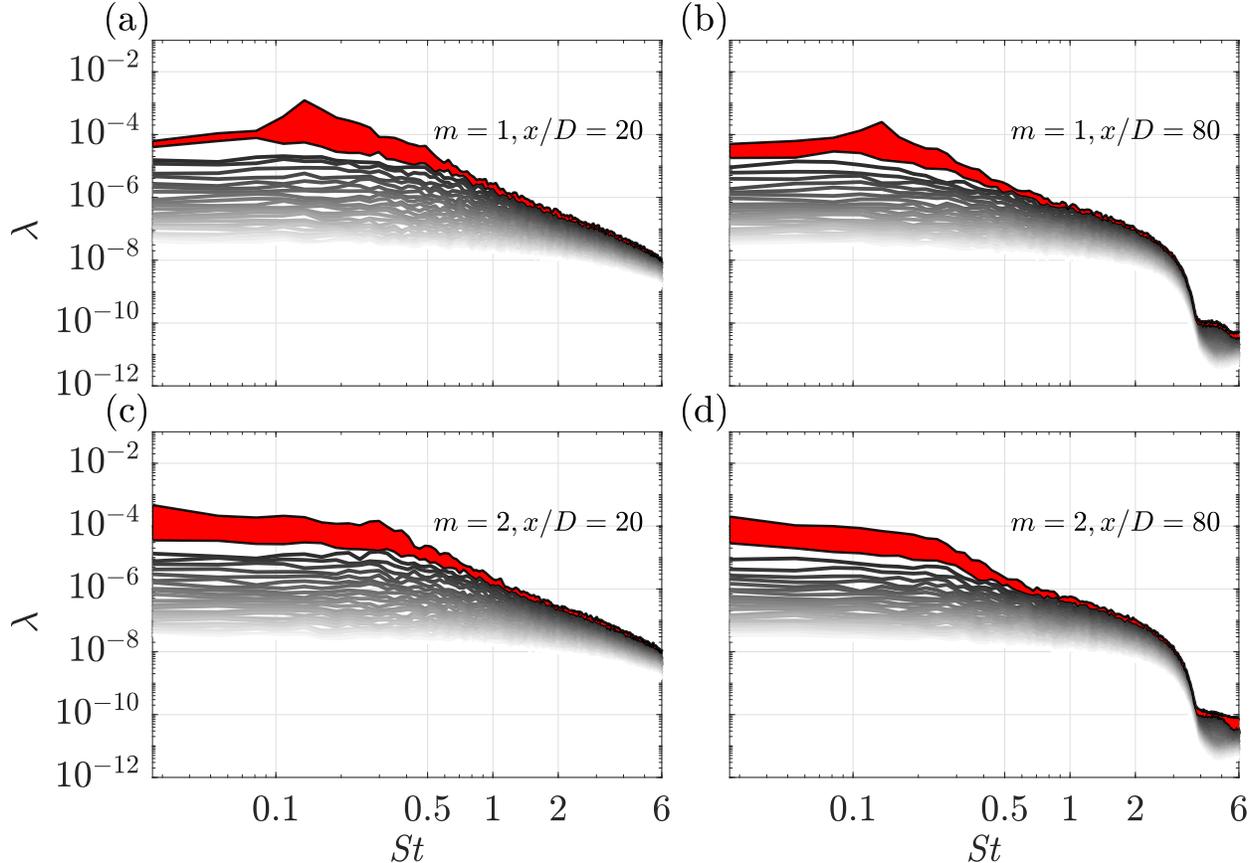}
\caption{SPOD eigenspectra of $25$ modes (dark to light shade corresponds to high to low energy eigenvalues): (a) $m=1$, $x/D = 20$; (b) $m=1$, $x/D = 80$; (c) $m=2$, $x/D = 20$; (d) $m=2$, $x/D = 80$.} 
\label{fig:eigenspectra}
\end{figure}

As observed in Fig. \ref{fig:bar_plots_m}, azimuthal modes $m=1$ and $m=2$ dominate the energy distribution of the leading SPOD mode in the near as well as far wake. To further clarify the energy distribution among  different frequencies for these two azimuthal modes, Fig. \ref{fig:eigenspectra} shows the SPOD eigenspectra of the $m=1 $ and $m=2$ modes at two representative locations in the near ($x/D = 20$) and far ($x/D =80$) wake. The area shaded by red denotes the difference between the energy content of the first and second SPOD modes. For the $m=1$ mode in Fig. \ref{fig:eigenspectra}(a) and (b), the distinct peak at $\Str=0.135$ is still clearly visible in the leading SPOD mode at $x/D = 80$. This is in contrast to the results of Johansson and George \cite{johansson_far_2006-1} in which the VS structure was almost undetectable in the eigenvalue spectra by $x/D =70$ (see Fig. 2 of that paper). Interestingly the  peak at $\Str = 0.135$ is not visible in the subsequent SPOD modes. Besides, there is a significant gap between the first and second SPOD modes at $\Str \approx 0.135$, more so at $x/D = 20$ than at $x/D = 80$. This large gap implies that vortex  shedding contributes significantly to the dynamics of the overall behavior of the  $m=1$ mode. 

Contrary to the $m=1$ mode, the eigenspectra of the $m=2$ mode shown in Fig. \ref{fig:eigenspectra}(c) and (d) peaks near $\Str \rightarrow 0$ and decays monotonically with increasing $\Str$. This decay rate is observed to increase for frequencies with  $\Str >  0.5$ at both the locations. Like the $m=1$ mode, the $m=2$ mode also exhibits a prominent gap between $\lambda^{(1)}$ and $\lambda^{(2)}$ at low frequencies with $\Str < 0.3$. SPOD eigenspectra of $m=1$ and $2$ analyzed at other locations (not shown here) qualitatively exhibit features similar to the locations shown in Fig. \ref{fig:eigenspectra}.

\begin{figure}
\centering
\includegraphics[width=0.75\linewidth]{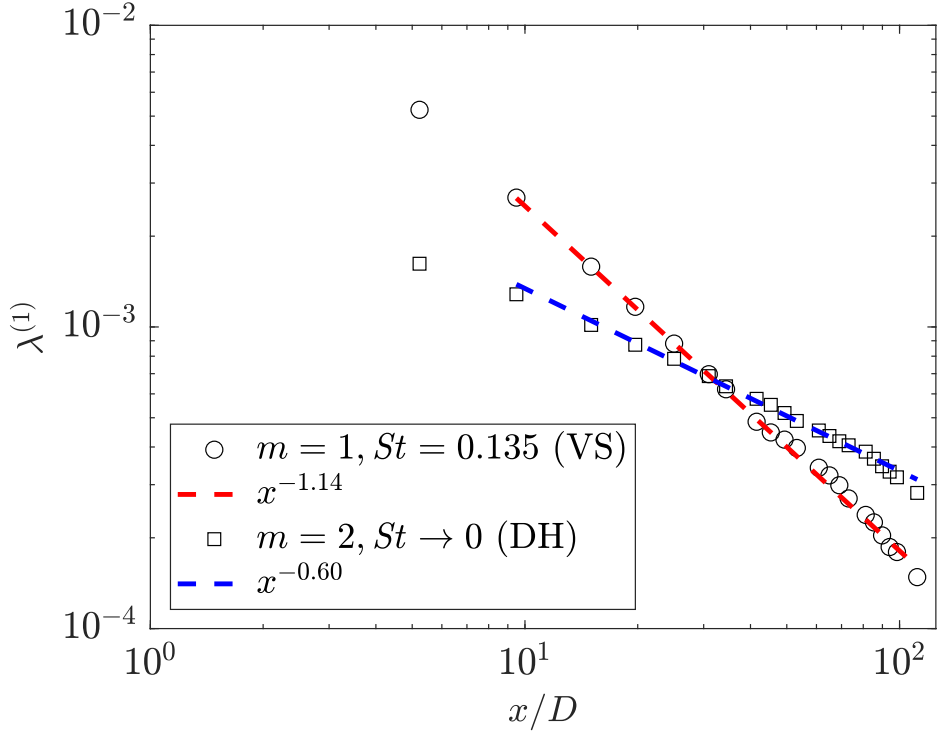}
\caption{Evolution of leading SPOD mode of $m=1$, $\Str=0.135$ and $m=2$, $\Str \rightarrow 0$ as $x/D$.}
\label{fig:eigenvalues_decay}
\end{figure}

Figure \ref{fig:contour_maps_eigenvalues} established that the energy in the leading SPOD mode is dominated by the DH structure ($m =2$ and $St \rightarrow 0$) in the far wake. To further quantify this observation, the evolution of $\lambda^{(1)}$ of the VS and DH structure is plotted in Fig. \ref{fig:eigenvalues_decay}. Both SPOD modes exhibit a monotonic decay which is in accordance with the decaying nature of wake turbulence. However, there are salient differences in the nature of their decay. The leading SPOD mode of the VS structure decays as $\lambda^{(1)} 
 \propto x^{-1.14}$ from $10 <x/D < 120$. On the other hand, $\lambda^{(1)}$ of the DH structure decays at a slower rate as $\lambda^{(1)} \propto x^{-0.60}$ so that it eventually exceeds the VS mode in terms of energy content beyond $x/D = 35$. 


\begin{figure}
\centering
\includegraphics[width=\linewidth]{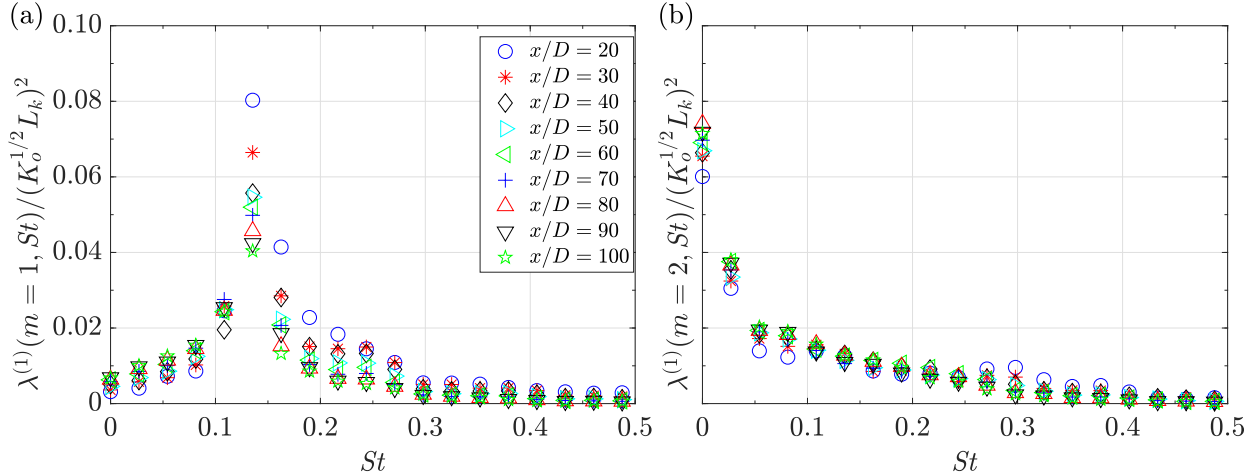}
\caption{$\lambda^{(1)}$ of: (a) $m=1$ and (b) $m=2$, scaled by $(K_{o}^{1/2}L_{k})^{2}$ for $20 < x/D < 100$. Here $K_{o}$ is the centerline value of TKE and $L_{k}$ is the TKE-based wake width.}
\label{fig:fig_scaling_eigenvalues}
\end{figure}

 We now explore the self-similarity of $\lambda^{(1)}$ of the $m=1$ and $m=2$ modes.  Figure \ref{fig:fig_scaling_eigenvalues} shows $\lambda^{(1)}$ for these two modes after scaling by $(K_{o}^{1/2} L_{k})^{2}$, a quantity representative of area-integrated TKE in the wake. For $m=2$, the eigenvalues collapse well when scaled by $(K_{o}^{1/2}L_{k})^{2}$ throughout $20 < x/D < 100$ as seen in Fig. \ref{fig:fig_scaling_eigenvalues}(b). The unscaled eigenvalues (not presented here) for $m=2$  show a variability of $50\%-60\%$ for lower frequencies. The eigenspectra of $\lambda^{(1)}$ for $m=2$ always peaks at $\Str \rightarrow 0$ for all downstream locations. The local timescale ($\zeta$) of an axisymmetric wake scales as $x^{3m/2}$ if we assume: (i) $U_{d} \propto x^{-m}$, and (ii) $\zeta \sim L_{d}/U_{d}$. Thus, the local frequency $f \propto x^{-3m/2}$ decays as $x/D$ increases since $m$ is a positive real number. The conclusion that $f$ decays with $x/D$ is unchanged even if local turbulent velocity ($K_{o}^{1/2}$) and TKE-based wake width ($L_{k}$) is used to form $\zeta$. For the present case, $f$ starts off as $f \sim O(10^{-2})$ and decays to $f \sim O(10^{-3})$ by the end of the domain. The collapse of the  $m=2$ energy content by local shear variables, particularly in the limit of $\Str \rightarrow 0$,
 suggests a possible link of this mode to the local shear structure of the wake.
 
Figure \ref{fig:fig_scaling_eigenvalues}(a) shows the eigenspectrum of the $m=1$ mode scaled with $(K_{o}^{1/2}L_{k})^{2}$ for $20 < x/D < 100$. There is a significant spread in the scaled eigenvalues around the vortex shedding frequency $\Str = 0.135$. It is clear from the plot that the leading SPOD mode of the vortex shedding structure does not collapse in local shear variables. This is a global mode which originates near the disk and convects downstream. As $\Str$ increases beyond $0.3$, the collapse improves indicating that the high frequency components in $m=1$ might be linked to the local turbulence structure. However their energy is small and hence the non-similar contribution of the vortex shedding frequency dominates the overall behavior of $m=1$.

\begin{figure}
\centering
\includegraphics[width=\linewidth]{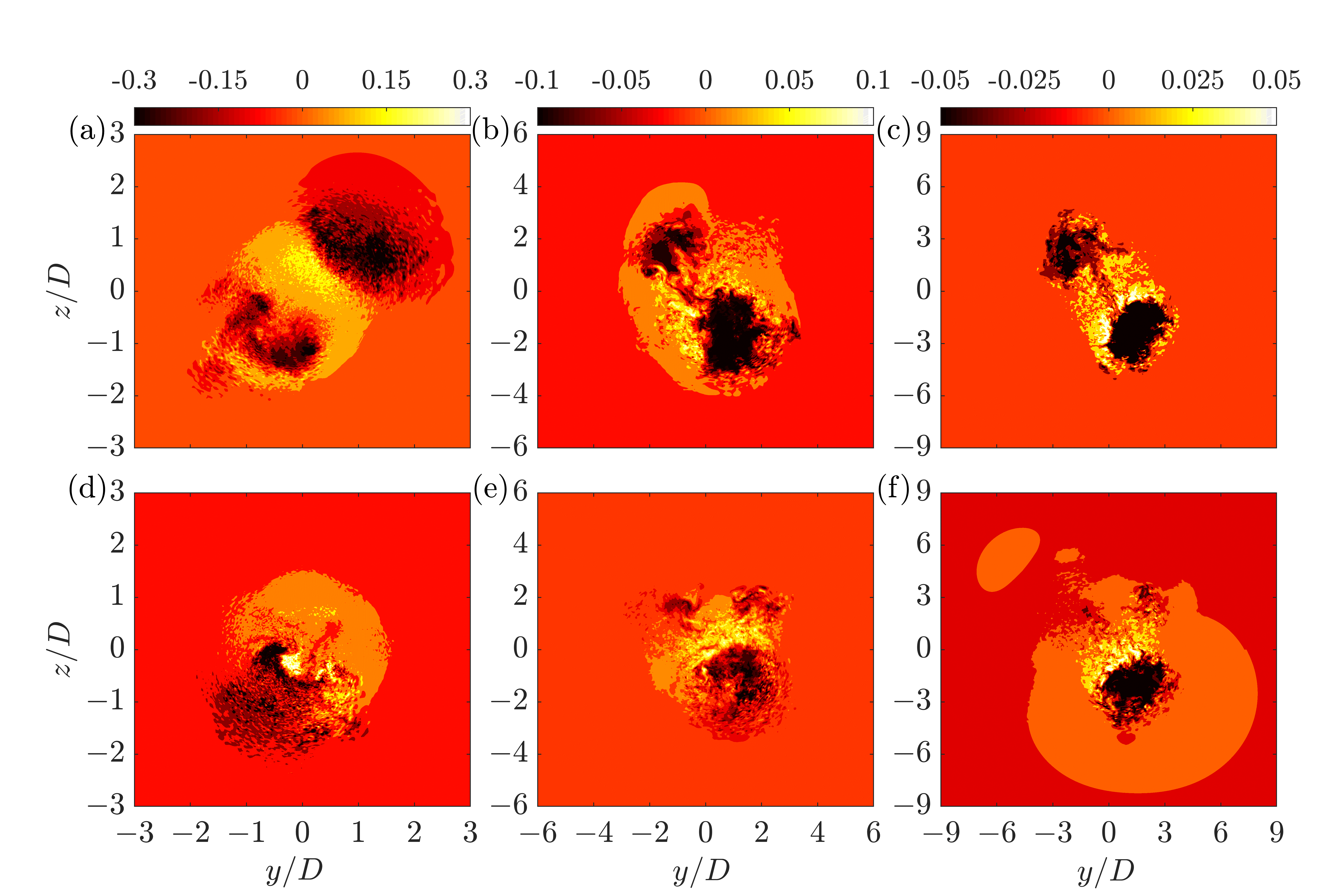}
\caption{Instantaneous snapshots of the  $u'_{x}$ field showing the imprint of  the $m=1$ (bottom row) and $m=2$ (top row) modes: (a,d)  at  $x/D = 10$,  (b, e) at $x/D = 40$ , and  (c, f) at $x/D = 80$.}
\label{fig:instant_realization}
\end{figure}

 POD is a statistical technique. Thus, although the obtained mode optimally capture the fluctuation energy in an ensemble-averaged sense, these modes do not necessarily represent the structures of instantaneous eddies in the flow. However, it is the case that these modes possess the imprints of coherent structures found in instantaneous snapshots. To assess whether different azimuthal modes which are found to be dominant from SPOD analysis are distinctly visible in the flow field, $u'_{x}$ at three different downstream locations $x/D  =10, 40$, and $80$ and at some selected time instants is plotted in Fig. \ref{fig:instant_realization}. These snapshots were selected by projecting instantaneous $u'_{x}$ to leading SPOD modes of the VS and DH structures and requiring large values of projection coefficient (similar to the approach of Hellstrom et al. \cite{hellstrom_self-similarity_2016}). In the top row (Fig. \ref{fig:instant_realization}(a,b,c)), the instantaneous $u'_{x}$ has the imprint of the $m=2$ velocity field at all three locations. Likewise the bottom row shows the time instants at which the velocity field exhibits evidence of the $m=1$ mode. Although these snapshots do not exactly mimic the mode shapes (inset contour maps of Fig. \ref{fig:scaling_m2m1_eigenmode}) to be discussed later, they imply that aspects of the $m=1$ and $2$ modes can be found in individual flow realizations. Both of these azimuthal modes are observed in the instantaneous flow snapshots throughout the wake evolution from near ($x/D = 10$) to far wake ($x/D =80$) consistent with  the eigenspectra analysis. 

\begin{figure}
\centering
\includegraphics[width=\linewidth]{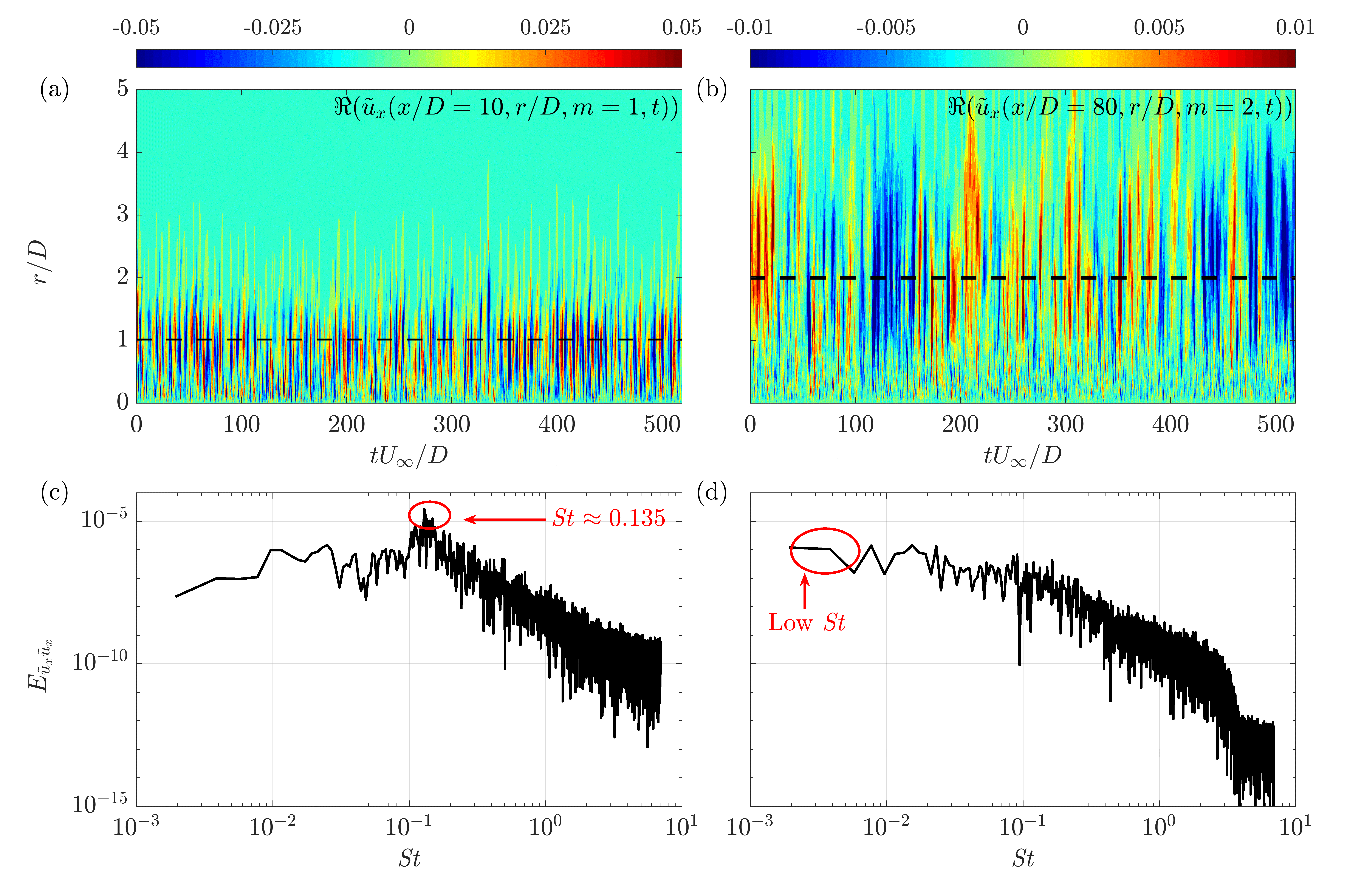}
\caption{$r$-$t$ plot of the real part of the azimuthally decomposed velocity field: (a) $m=1$ at $x/D = 10$, and (b) $m=2$ at $x/D = 80$. Power spectra of time series: (c) $r/D = 1$, $x/D = 10$ for $m =1$,   and (d) $r/D = 2$, $x/D = 80$ for $m=2$, with the $r/D$  locations shown by dashed black lines in (a) and (b), respectively.}
\label{fig:fft_mode_x_D_10_80}
\end{figure}

The  two most dominant SPOD frequencies, $\Str = 0.135$ and $\Str \rightarrow 0$, are apparent in the space-time history  of the $m=1$ and $m=2$  modes as demonstrated by the $r$ - $t$ plot  of these modes of $u'_{x}$  in Fig. \ref{fig:fft_mode_x_D_10_80}. The near-wake location of $x/D = 10$ (Fig. \ref{fig:fft_mode_x_D_10_80} a) has a clear periodicity  in the time series which  corresponds to a peak at $\Str \approx 0.135$ of the power spectrum of this signal at $r/D = 1$ (Fig. \ref{fig:fft_mode_x_D_10_80} b). The SPOD spectrum,  by exploiting correlation in space along with time,  makes  this frequency for $m=1$ more distinctive as was seen in Fig.  \ref{fig:eigenspectra}(a). While discussing Fig. \ref{fig:contour_maps_eigenvalues}, it was noted that the $\Str = 0$ peak in the SPOD spectrum of $m=2$ is likely related to a  very low-frequency signal. Figure \ref{fig:fft_mode_x_D_10_80}(b) substantiates this hypothesis since a signal with a very large time period, shown by wide (in the $t-$axis) patches of blue and red, can be discerned. These patches span the entire radial extent  of the wake. The power spectrum calculated at $r/D = 2$ (Fig. \ref{fig:fft_mode_x_D_10_80} d) indeed peaks at low $\Str \approx 10^{-3}$. 

\subsection{Eigenspectra of $m=0, 3,$ and $4$ modes} \label{m034_analysis}

\begin{figure}
\centering
\includegraphics[width=\linewidth]{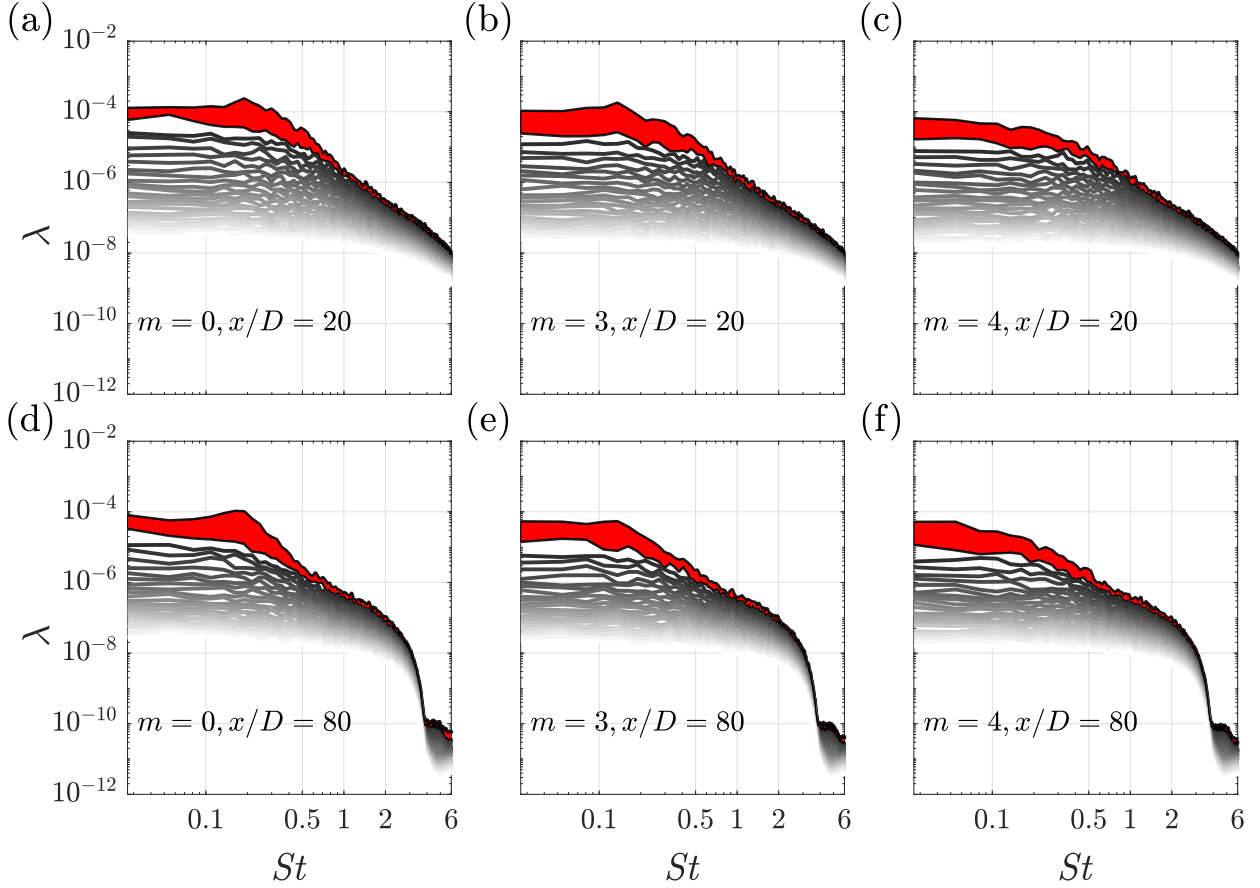}
\caption{SPOD eigenspectra of $25$ modes (dark to light shade corresponds to high to low energy eigenvalues) for $m =0$ (left), $m =3$ (middle) and $m = 4$ (right). Top row shows $x/D = 20$ and  bottom row shows $x/D = 80$.} 
\label{fig:eigenspectra_m034}
\end{figure}

The $m=0, 3,$ and $4$ modes are energetically  the three important azimuthal modes after the $m=1$ and $2$ modes. Therefore, it is of interest to characterize their eigenspectra and assess the applicability of local similarity scaling of their magnitude.

 Fig. \ref{fig:eigenspectra_m034} shows  eigenspectra of the $m = 0, 3$, and $4$ modes at two locations $x/D = 20$ and $x/D = 80$. Similar to the $m=1$ and $2$ modes, these azimuthal modes also exhibit a significant gap between $\lambda^{(1)}$ and $\lambda^{(2)}$ SPOD modes for $\Str < 0.5$ shown by the red-shaded area. The eigenspectra of the $m=0$ mode shown in Fig. \ref{fig:eigenspectra_m034}(a) and (d) exhibit a peak at $\Str = 0.189$ for $\lambda^{(1)}$. This peak is evident even in the far wake location at $x/D = 80$. This peak is also found close to the disk as  will be discussed in more detail in Section \ref{spod_nearbody}. The other two azimuthal modes, $m=3$ and $m=4$, exhibit features similar to the $m=1$ and $m=2$ modes, respectively.  The eigenspectrum of $\lambda^{(1)}$ for the $m=3$ mode shows a peak at the vortex shedding frequency $\Str = 0.135$, although the peak  is much less pronounced than for the $m=1$ case. Like the $m=2$ mode, the eigenspectra of the $m=4$ mode peak at $\Str \rightarrow 0$ and decay thereafter. An increased rate of decay with increasing frequency is also observed beyond $\Str = 0.3$, similar to the $m=2$ case. 

\begin{figure}
\centering
\includegraphics[width=\linewidth]{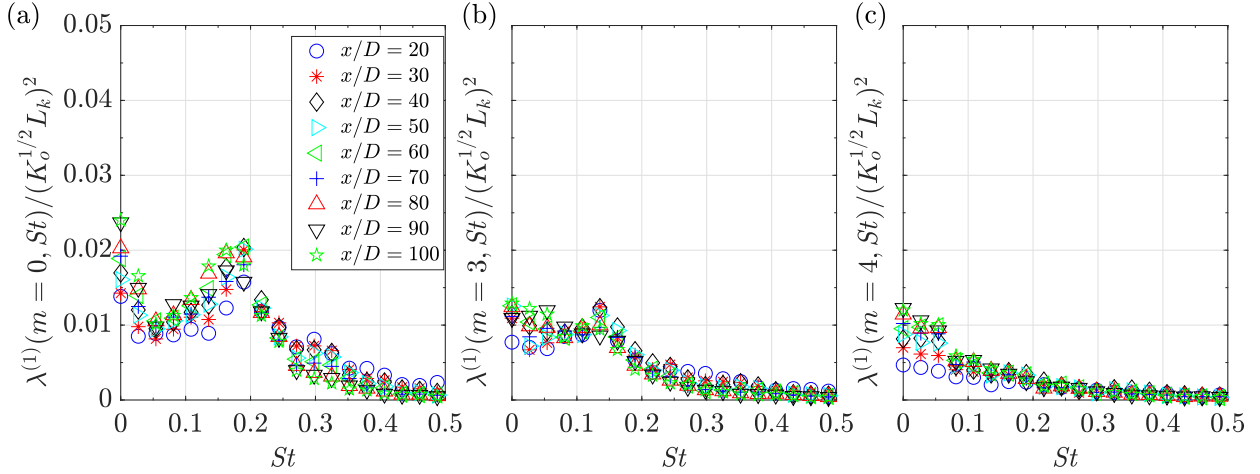}
\caption{Leading SPOD eigenvalue ($\lambda^{(1)}$), scaled by $(K_{o}^{1/2}L_{k})^{2}$, are plotted over $20 < x/D < 100$ for the following  modes: (a) $m=0$, (b) $m=3$, and (c) $m=4$,  Here $K_{o}$ is the centerline value of TKE and $L_{k}$ is the TKE-based wake width.}
\label{fig:fig_scaling_eigenvalues_m034}
\end{figure}

Figure \ref{fig:fig_scaling_eigenvalues_m034} shows the eigenspectra of $\lambda^{(1)}$ for the $m=0, 3,$ and $4$ modes spanning $20 \leq x/D \leq 100$ and scaled by $(K_{o}^{1/2}L_{k})^{2}$, to explore the presence of similarity as was done with $m =1$ and 2 in Fig. \ref{fig:fig_scaling_eigenvalues}.  There are two distinct peaks, $\Str = 0$ and $0.189$, in the scaled eigenspectra of the $m=0$ mode presented in Fig. \ref{fig:fig_scaling_eigenvalues_m034}. The scaled eigenspectra show significant spread for $\Str \leq 0.2$ indicating that the low-frequency content in the $m=0$ mode might have its origin near the  wake generator rather than being local. 
A  proper collapse of scaled eigenvalues is observed only beyond $\Str > 0.35$ for the $m=0$ azimuthal mode. As discussed in the context of Fig. \ref{fig:eigenspectra_m034}(b) and (e), the scaled eigenspectra of $m=3$ plotted in Fig. \ref{fig:fig_scaling_eigenvalues_m034}(b) show a peak at the vortex shedding frequency $\Str = 0.135$, the magnitude of which, relative to $(K_{o}^{1/2}L_{k})^{2}$, decays with increasing $x/D$. Somewhat similar to the scaled eigenspectra plot of the $m=0$ mode in Fig. \ref{fig:fig_scaling_eigenvalues_m034}(a), the scaled eigenspectra of the $m=1$ mode also show some spread for lower frequencies $\Str < 0.2$ and collapse only beyond $\Str \approx 0.2$.  Finally the scaled eigenspectra of the $m=4$ mode peak at $\Str \rightarrow 0$ and collapse well for $\Str > 0.1$. For $\Str < 0.1$, after $x/D \approx 50$,  the scaled eigenvalues do collapse. The scaled eigenspectra of the $m=4$ mode are quite similar to that of the $m=2$ mode, apart from the lower magnitudes. 

The $m=0, 3,$ and $4$ modes, although suboptimal relative to  the $m=1$ and $2$ modes dominate over the remaining modes in terms of energy content.   Based on the findings of Fig. \ref{fig:eigenspectra_m034} and \ref{fig:fig_scaling_eigenvalues_m034}, it can be concluded that the wake generator (disk in the present case) can have a profound impact on the characteristics of the suboptimal modes too which  can last for large downstream distances, at least up to $O(x/D = 100)$.


\section{Eigenmodes of the dominant vortex shedding  and double helix modes
}
 \label{spod_eigenmodes}

The shape of the eigenmodes for each velocity component is contrasted between the dominant VS and DH modes in this section. The applicability of similarity scaling to these modes is also assessed.

\begin{figure}
\centering
\includegraphics[width=\linewidth]{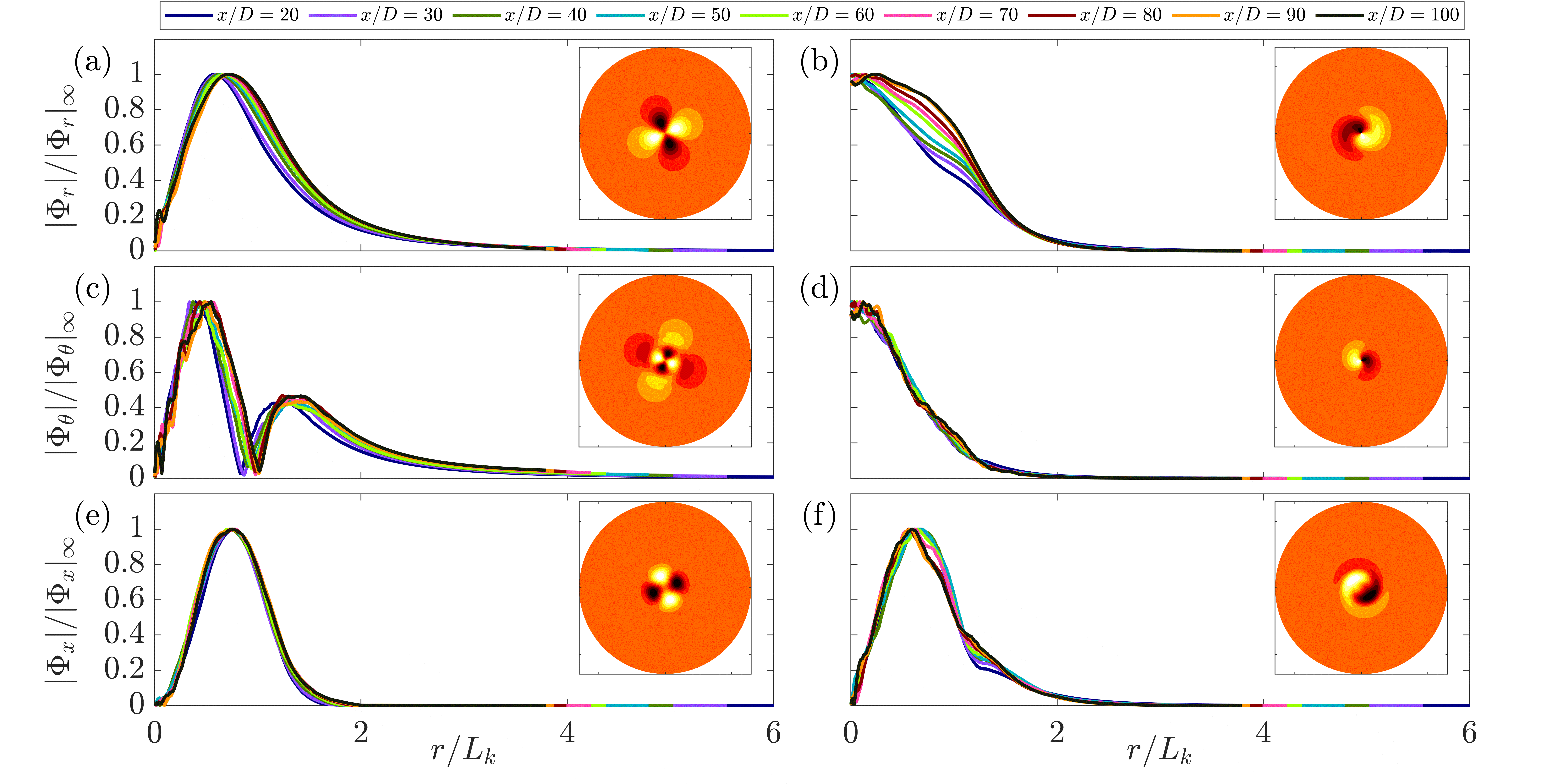}
\caption{Modulus of eigenmode shapes for all three velocity components corresponding to $\lambda^{(1)}$ of the DH and VS structures: (a), (c), (e) correspond to $u_{r}$, $u_{\theta}$, $u_{x}$ eigenmodes respectively of the DH structure and (b), (d), (f) correspond to $u_{r}$, $u_{\theta}$, $u_{x}$ eigenmodes respectively of the VS structure. Inset figures show the two-dimensional structure of real part of corresponding eigenmodes at $x/D = 50$.}
\label{fig:scaling_m2m1_eigenmode}
\end{figure}

Figure \ref{fig:scaling_m2m1_eigenmode} shows the shapes of leading SPOD modes of the VS and DH structures. For this purpose, moduli of eigenmodes scaled by their respective maximum values $(|\Phi_{i}^{(1)}(r)|/|\Phi_{i}^{(1)}(r)|_{max})$ are plotted as a function of the radial similarity coordinate.
Figure \ref{fig:scaling_m2m1_eigenmode} (a), (c), and (e) show the 
$u_r$, $u_\theta$, and $u_x$ components
 of the leading SPOD eigenmode of the DH structure at different $x/D$ locations. The 
 $u_r$ and $u_x$ modes have single global peak at nearby locations, $\Phi_{r}^{(1)}$ at at $r/L_{k} \approx 0.7$  and $\Phi_{x}^{(1)}$ at $r/L_{k} \approx 0.75$, throughout $20 < x/D < 100$. It will be shown later that their cross-correlation $\langle -u'_{x}u'_{r}\rangle$ also peaks at $r/L_{k} \approx 0.7$.
 The radial shape of  $\Phi^{(1)}_{x}$ starts exhibiting self-similarity from $x/D = 20$ onward while $\Phi_{r}^{(1)}$ exhibits collapse  beyond $x/D = 40$. Furthermore, for the DH mode,  $\Phi_{x}^{(1)}$ decays faster with increasing $r$ relative to $\Phi_{r}^{(1)}$. The shape of  the $\Phi_{\theta}^{(1)}$ mode is qualitatively different with respect to its counterparts for  radial and axial velocity. At all downstream locations, $|\Phi_{\theta}^{(1)}|$ exhibits two maxima, at $r/L_{k} \approx 0.50$ and $1.35$,  respectively, and a minimum  at $r/L_{k} \approx 1$. The minimum in the plot of $|\Phi_{\theta}^{(1)}|$ is evident as  a zero-crossing at $r/L_{k} \approx 1$ for $\Phi_{\theta}^{(1)}$ in  the two-dimensional inset plot for this mode in Fig. \ref{fig:scaling_m2m1_eigenmode} (c). 
 Similar to the other two components, $\Phi_{\theta}^{(1)}$ eventually becomes self-similar, beyond $x/D \approx 30$. Fig. \ref{fig:scaling_m2m1_eigenmode} suggests that the leading SPOD mode of the DH structure eventually becomes self-similar for all three velocity components beyond $x/D \approx 40$.

The inset figures in Fig. \ref{fig:scaling_m2m1_eigenmode}(a), (c), and (e) show the two-dimensional contour plots of the real part of the corresponding velocity components of the leading SPOD mode of the DH structure at $x/D  =50$. The characteristic 4-lobe structure of the $m=2$ mode can be seen in these contour maps. From the contour maps, it can be inferred that $\Phi_{r}^{(1)}$ and $\Phi_{x}^{(1)}$ are negatively correlated  at $x/D = 50$. It turns out that the imaginary parts of radial and axial components are also negatively correlated (not shown here) resulting in an overall positive contribution from the leading SPOD mode of the  DH structure to $\langle -u'_{x}u'_{r} \rangle$, the most dominant Reynolds shear stress term in axisymmetric turbulent shear flows. This positive contribution to $\langle -u'_{x}u'_{r} \rangle$ is found  at all other downstream locations sampled for SPOD analysis in the present study and will be discussed  with more detail in a later  section. 

Figure \ref{fig:scaling_m2m1_eigenmode}(b), (d), and (f) show the moduli of the leading SPOD mode of the VS structure. Except $\Phi_{x}(r)$, none of the other two components start off at zero near $r=0$. Both $|\Phi_{r}(r)|$ and $|\Phi_{\theta}(r)|$ peak near the axis and decay monotonically to zero as $r$ increases. On the other hand, $|\Phi_{x}(r)|$ peaks at $r/L_{k} \approx 0.66$. All three components of $\Phi_{i}^{(1)}$ die out to zero by $r/L_{k} \approx 2$ implying that the prevalence of vortex shedding is confined to $r/L_{k} < 2$ at all locations. An immediate observation can be made about the lack of self-similarity of the leading SPOD mode of the VS structure. Although $\Phi_{x}^{(1)}$ and $\Phi_{\theta}^{(1)}$ tend to collapse to some extent when $r$ is scaled by $L_{k}$, there is a significant spread in $\Phi_{r}^{(1)}$ mode for all downstream locations. 
 
The inset plots in Fig. \ref{fig:scaling_m2m1_eigenmode}(b), (d), and (f) show the two-dimensional contours of the real part of the corresponding SPOD velocity components at $x/D  =50$. Similar to the leading SPOD mode of the DH structure, the  real parts of $\Phi_{r}^{(1)}$ and $\Phi_{x}^{(1)}$ are negatively correlated for the VS structure too. The imaginary parts although not explicitly shown are also negatively correlated. This implies that the VS structure also contributes positively to $\langle -u'_{x}u'_{r} \rangle$ in a similar fashion to the DH structure.

\section{Reconstruction of the TKE and Reynolds shear stress using SPOD modes} \label{tke_uxur_recon}

In this section, the utility of SPOD modes for capturing the spatial distribution of  TKE and $\langle u'_{x}u'_{r} \rangle$ is examined.
By construction, SPOD modes in the present analysis optimally capture  the area-integrated TKE at a given cross-section (refer section \ref{spod_overview}). However there is no guarantee that these modes will be optimal for the Reynolds shear stress $\langle u'_{x}u'_{r}\rangle$. Nevertheless it is generally the case that the energetic structures are also the ones that carry a major portion of the turbulent shear stress. In fact, as will become clear for the present example of a turbulent wake, SPOD is successful as a low-order model for the Reynolds shear stress, more so than for the TKE.

The Reynolds  stress tensor can be reconstructed from a selected number ($n = 1$ to $\Lambda$)  of SPOD modes as follows:
\begin{equation}
\langle u'_{i}u'_{j} \rangle (x;r)  = \sum\limits_{n=1}^{\Lambda}\sum\limits_{m=-M}^{m=M} \, \sum\limits_{\Str = -N}^{\Str = N} \lambda^{(n)}(x;m,\Str) \, \Phi_{i}^{(n)}(x;r,m,\Str) \, \Phi_{j}^{(n)*}(x;r,m,\Str),
\end{equation}
where the first $M$ azimuthal modes and  the  first $N$ discrete frequencies are incorporated in the reconstruction. Setting $i =x$ and $j = r$ gives the reconstructed $\langle u'_{x}u'_{r} \rangle$ and twice the TKE is recovered when $i=j$ adopting the convention of summation over repeated indices. The reconstruction of TKE and $\langle u'_{x}u'_{r} \rangle$ is 
elaborated as follows.
\subsection{TKE reconstruction from SPOD modes}
\label{tke_recon}
\begin{figure}
\centering
\includegraphics[width=\linewidth]{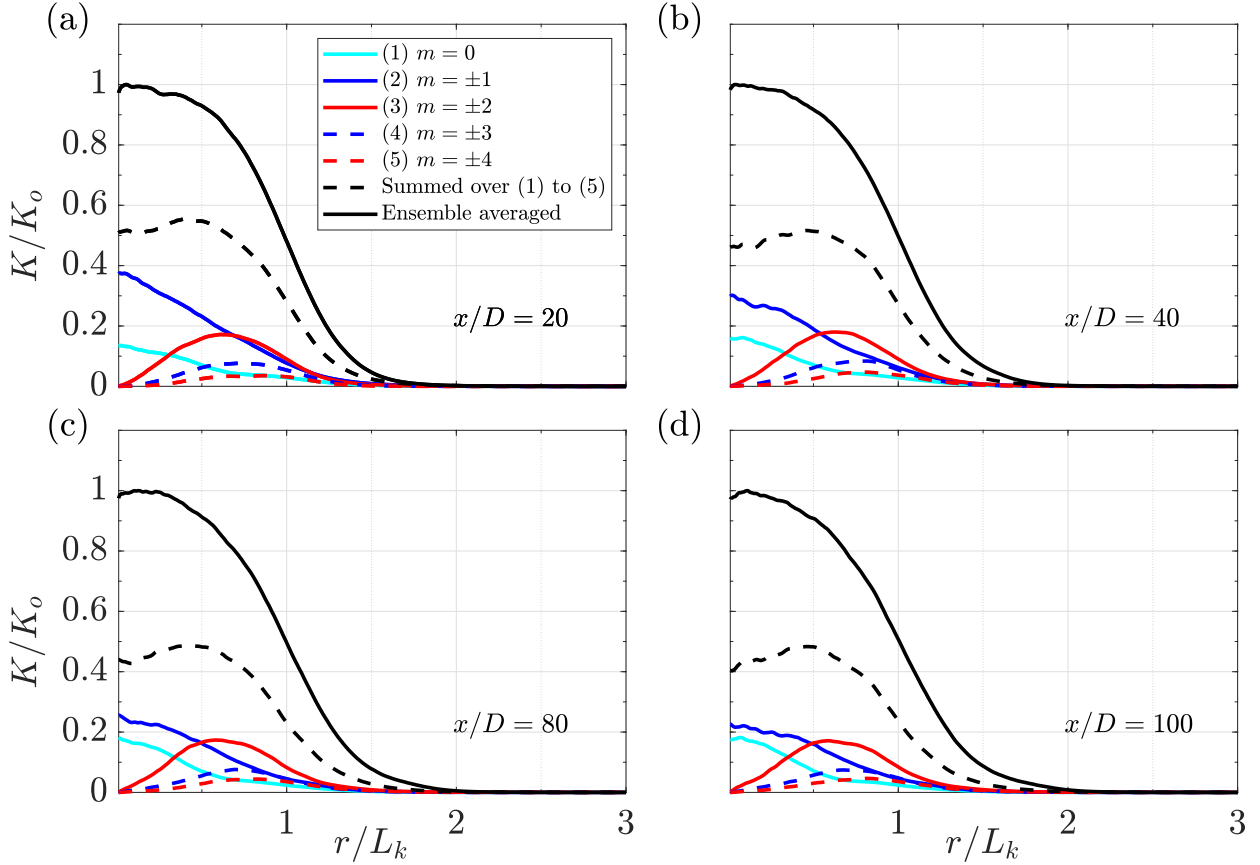}
\caption{Reconstruction of TKE from a low-order truncation that comprises the leading 3 SPOD modes of azimuthal modes $(|m| \leq 4)$ with energy summed over $-1 \leq\Str \leq 1$: (a) $x/D = 20$, (b) $x/D = 40$, (c) $x/D = 80$, and (d) $x/D = 100$. The radial direction is scaled with TKE-based wake width $L_{k}$ and TKE is scaled with centerline TKE $K_{o}$.}
\label{fig:k_reconstruction}
\end{figure}

Figure \ref{fig:k_reconstruction} shows the reconstructed TKE using a  low-order truncation that includes modes with the following characteristics: $|m| \leq 4$, $|\Str| \leq 1$, and $n \leq 3$. The fidelity of the reconstruction is shown at four different locations $x/D = 20, 40, 80$, and $100$ by comparison with the actual TKE obtained as an ensemble average of the numerical data. For the given set of $(m, \Str, n)$ triplets, the qualitative nature of reconstructed TKE remains similar throughout all the locations considered in Fig. \ref{fig:k_reconstruction}.  However, a closer inspection reveals that the TKE reconstruction deteriorates slightly with increasing $x/D$. For instance, at $x/D = 20$, reconstructed TKE captures $50\%$ of the actual TKE at the centerline. By $x/D = 80$, this value comes down to $40\%$. It is worth noting that the quality of reconstruction improves with increasing $r/L_{k}$ as the flow becomes less turbulent away from the centerline and {fewer}  modes are required to accurately capture the TKE. The implications of expanding the range of $m, \Str,$ and $n$ in the reconstruction will be discussed shortly.

The major contributors to the overall reconstructed TKE in Figure \ref{fig:k_reconstruction}  are the $m=0, 1$, and $2$ azimuthal modes. However, only the  $m=0$ and $m=1$ modes contribute to the centerline TKE. All the other azimuthal modes with $m \geq 2$ have zero TKE at the centerline. It can also be seen that the relative contribution of the leading  $m=1$ mode to the centerline TKE reconstruction declines progressively with increasing $x/D$.
 This decrease is linked to the declining relative importance of the $m=1$ mode in the integrated eigenspectra as discussed in preceding sections. 

\begin{figure}
\centering
\includegraphics[width=\linewidth]{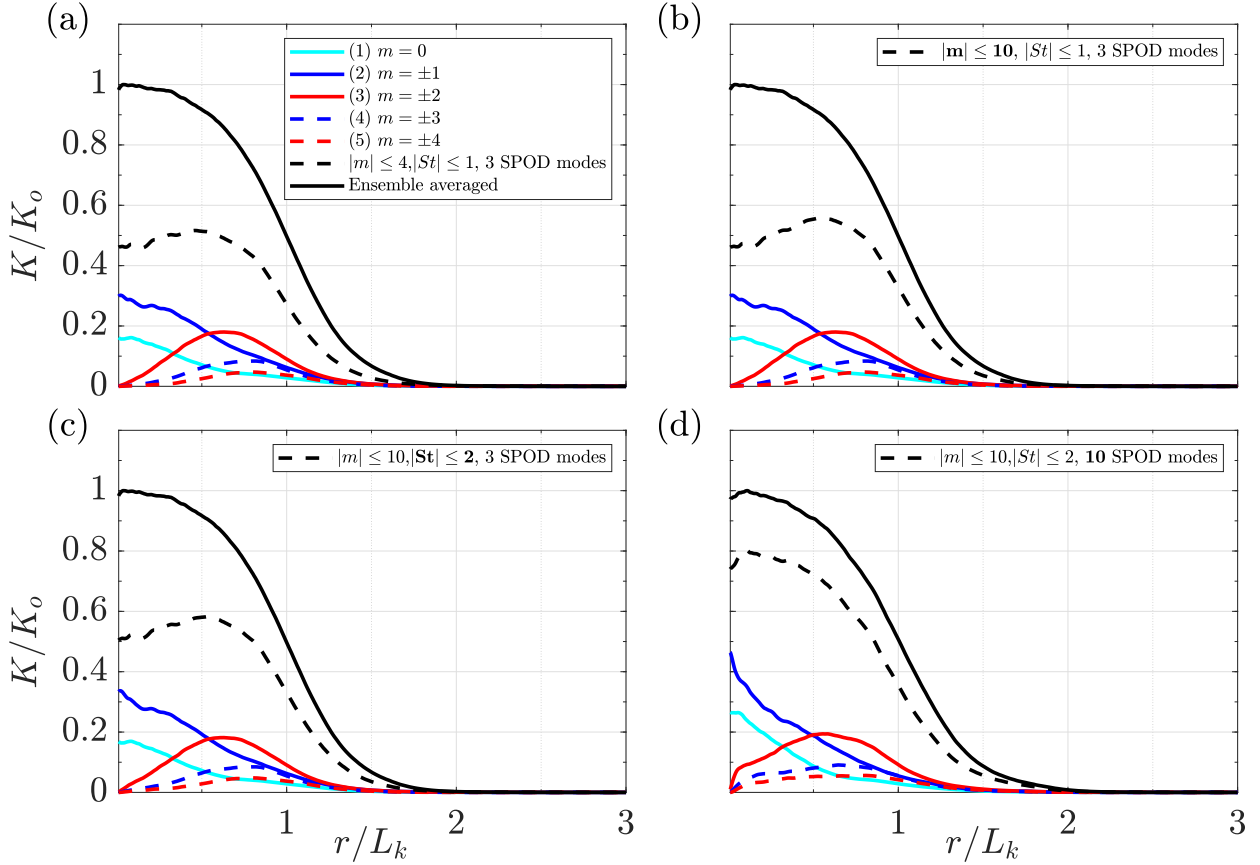}
\caption{Reconstruction of TKE at $x/D = 40$ using: (a) $|m| \leq 4$, $|\Str| \leq 1$, $n \leq 3$, (b) $|m| \leq 10$, $|\Str| \leq 1$, $n \leq 3$, (c) $|m| \leq 10$, $|\Str| \leq 2$, $n \leq 3$, and (d) $|m| \leq 10$, $|\Str| \leq 2$, $n \leq 10$. Radial direction is scaled with TKE-based wake width $L_{k}$ and TKE is scaled with centerline TKE $K_{o}$. In each subplot from (b) to (d), one parameter is changed relative to the preceding subplot and that parameter is boldfaced in the legend.}
\label{fig:k_reconstruction_analysis}
\end{figure}

The sensitivity of the TKE reconstruction to the addition of more modes ($m$, $\Str$, or $n$) has been investigated. The  effect of increasing the number of modes is illustrated at one representative location $x/D = 40$ for brevity. The trends do not change qualitatively at different $x/D$ except for a slight decrease in the energy capture at larger $x/D$ locations  pointing to an increasing importance of higher modes. From Fig. \ref{fig:k_reconstruction_analysis}(a) to (d), the upper limits of $m$, $\Str$, and $n$ are increased successively.  The reconstruction of TKE shows a monotonically increasing accuracy with the inclusion of additional $m$, $\Str$, and $n$.

Comparing Fig. \ref{fig:k_reconstruction_analysis}(b) with (a) one can see that the reconstructed TKE improves in the region of $r/L_{k} > 0.5$ when more azimuthal modes are included keeping $\Str$ and $n$ same. This indicates that the higher azimuthal modes (modes with increasing $m$) do not significantly influence the TKE near the centerline and are only active beyond a certain $r/L_{k}$. This characteristic is confirmed by the TKE profiles of solely  the $m=3$ and $m=4$ modes which start off as zero near centerline and peak at $r/L_{k} \approx 0.75$.  

When the reconstruction is performed using more frequencies, shown in Fig. \ref{fig:k_reconstruction_analysis}(c), there is a slight improvement in the overall reconstruction of the TKE. The individual contributions from each azimuthal modes increase leading to an overall improvement of the reconstructed profile 

The inclusion of more SPOD modes leads to  a significant boost in the quality of TKE reconstruction in the wake core ($r/L_{k} < 0.75$).  The reconstructed profile captures almost $80 \%$ of the  TKE at the centerline and in the wake core with 10 SPOD modes (Fig. \ref{fig:k_reconstruction_analysis} d) instead of  about $50\%$ with 3 SPOD modes (Fig. \ref{fig:k_reconstruction_analysis} c). 
The improvement away from the wake core is small and the reconstruction remains at about  $80\%$ of the actual value.
This implies that the higher SPOD modes that are individually suboptimal, when summed together, can contribute significantly to the  TKE in the wake core where the turbulence is more intense. Referring back to Fig. \ref{fig:eigenspectra}, these SPOD modes ($\lambda^{(3)}$ and beyond) are the ones which have almost uniform energy distribution over  $\Str < 0.5$ and do not contain evidence of any coherent structures. They are representative of  incoherent turbulence which is more prominent near the centerline.

\subsection{Reconstruction of Reynolds shear stress with SPOD modes}
\label{uxur_recon}

\begin{figure}
\centering
\includegraphics[width=\linewidth]{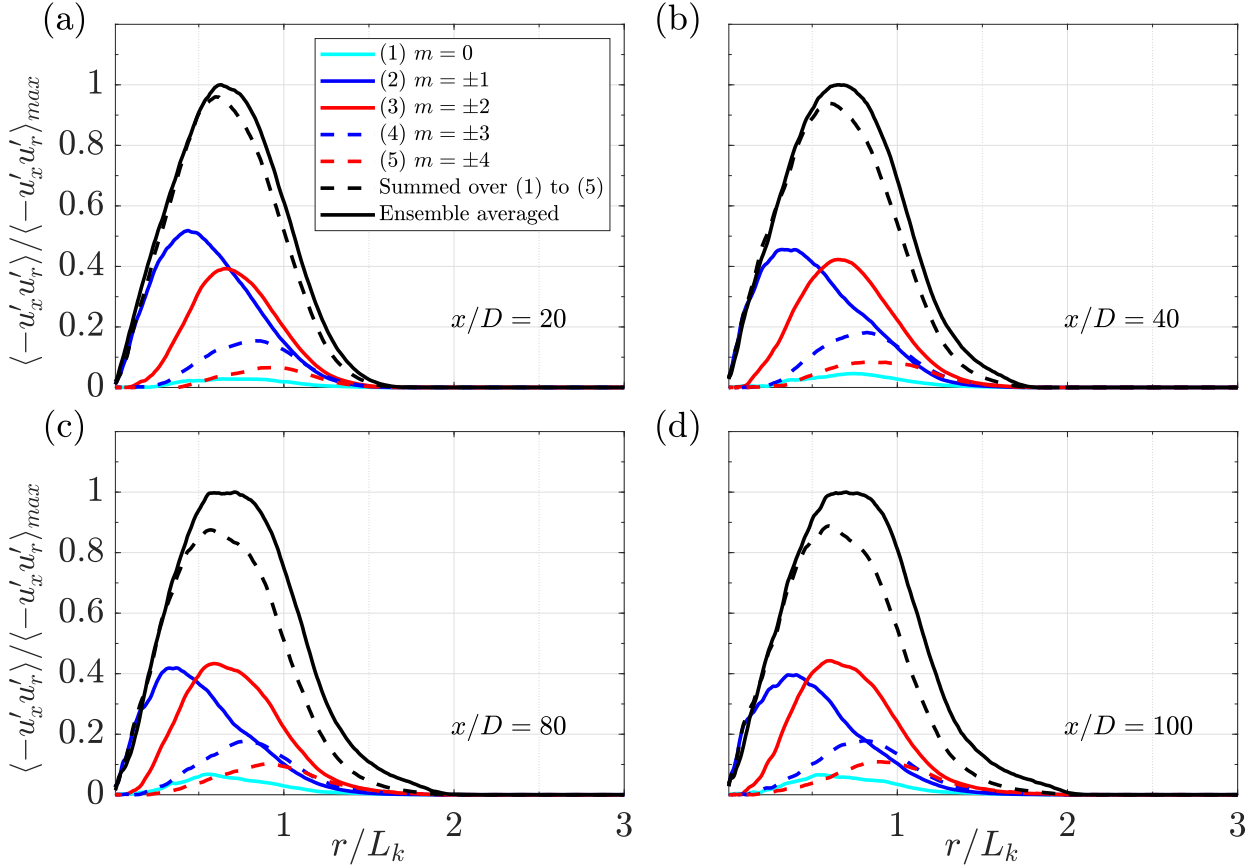}
\caption{Reconstruction of $\langle u'_{x}u'_{r} \rangle$ from the leading 3 SPOD modes of various azimuthal modes $(|m| \leq 4)$ summed over $-1 \leq\Str \leq 1$ at different streamwise locations: (a) $x/D = 20$, (b) $x/D = 40$, (c) $x/D = 80$, and (d) $x/D = 100$. Radial direction is scaled with TKE-based wake width $L_{k}$ and $\langle u'_{x}u'_{r} \rangle$ is scaled with $\langle -u'_{x}u'_{r} \rangle_{max}$.}
\label{fig:uxur_reconstruction}
\end{figure}

In the same vein as Fig. \ref{fig:k_reconstruction} for the TKE, Fig. \ref{fig:uxur_reconstruction} compares the reconstruction of $\langle u'_{x}u'_{r} \rangle$ with the corresponding actual value at four locations $x/D = 20, 40, 80, $ and $100$. 
The leading 3 SPOD modes of the azimuthal modes $|m| \leq 4$ and $|St| \leq 1$ are employed. 
The reconstruction turns out to have higher fidelity for  $\langle u'_{x}u'_{r} \rangle$   than for TKE.

Figure \ref{fig:uxur_reconstruction} shows that the two major contributors to the Reynolds stress $\langle u'_{x}u'_{r} \rangle$ are the $m=1$ and $m=2$ modes followed by the $m=3$ and $m=4$ modes, respectively. 
Consistent with the integrated eigenspectrum in Fig. \ref{fig:bar_plots_m}, the $m=1$ mode contributes more than $m=2$ to the reconstruction in the near wake, i.e., $x/D = 20$. With increasing $x/D$, the contribution from the $m=2$ mode gradually exceeds the $m=1$ contribution, except close to the centerline, say $r/L_k < 0.25$. The $m=0$ mode carries negligible shear stress although it carries significant TKE as seen in the previous subsection.
It is worth noting that the accuracy of reconstruction of $\langle u'_{x}u'_{r} \rangle$ is significantly better at $x/D = 20$ and $40$ 
 than at the far wake locations of $x/D = 80$ and $100$.
  With increasing $x/D$, more azimuthal modes have to be included for the far wake so as to get the same quality of reconstruction as in the near wake.

As pointed out, the two major contributors to $\langle u'_{x}u'_{r}\rangle$ are the azimuthal modes $m=1$ and $m=2$. Interestingly, these two azimuthal modes capture different characteristics of the radial variation of Reynolds shear  stress.
The $m=1$ mode accurately captures the  actual $\langle u'_{x}u'_{r} \rangle$ in the central region with $r/L_k < 0.25$ including the slope at the axis ($r/D = 0$). Thereafter, its contribution peaks between  $r/L_{k} = 0.25$ and $r/L_{k} = 0.5$, and decays faster than the contribution from the azimuthal mode $m=2$. Contributions from the $m=2$ and higher azimuthal modes start off with zero slope at $r/D = 0$ and they do not contribute near to the centerline. 

While the  $m=1$ mode dominates for  $r/L_{k} < 0.25$,  the $m=2$ mode plays an increasing important role in the reconstruction of $\langle u'_{x}u'_{r}\rangle$ at larger  $r/L_k$. For instance, the maximum of the $m=2$ contribution coincides with the peak of actual  $\langle u'_{x}u'_{r}\rangle$ at $r/L_{k} \approx 0.75$.

The profiles of $\langle u'_{x}u'_{r}\rangle$, which were  shown in Section \ref{uxur_k_real_data}, tend to flatten at the  peak location with increasing $x/D$. This flattening at  the peak is also observed for the $m=2$ contribution. Compared to the $m=1$ mode, the Reynolds stress in the $m=2$ mode decays slowly with $r$ and dominates over the $m=1$ contribution beyond $r/L_{k} > 0.5$ from $x/D = 40$ onward. 
 
\begin{figure}
\centering
\includegraphics[width=\linewidth]{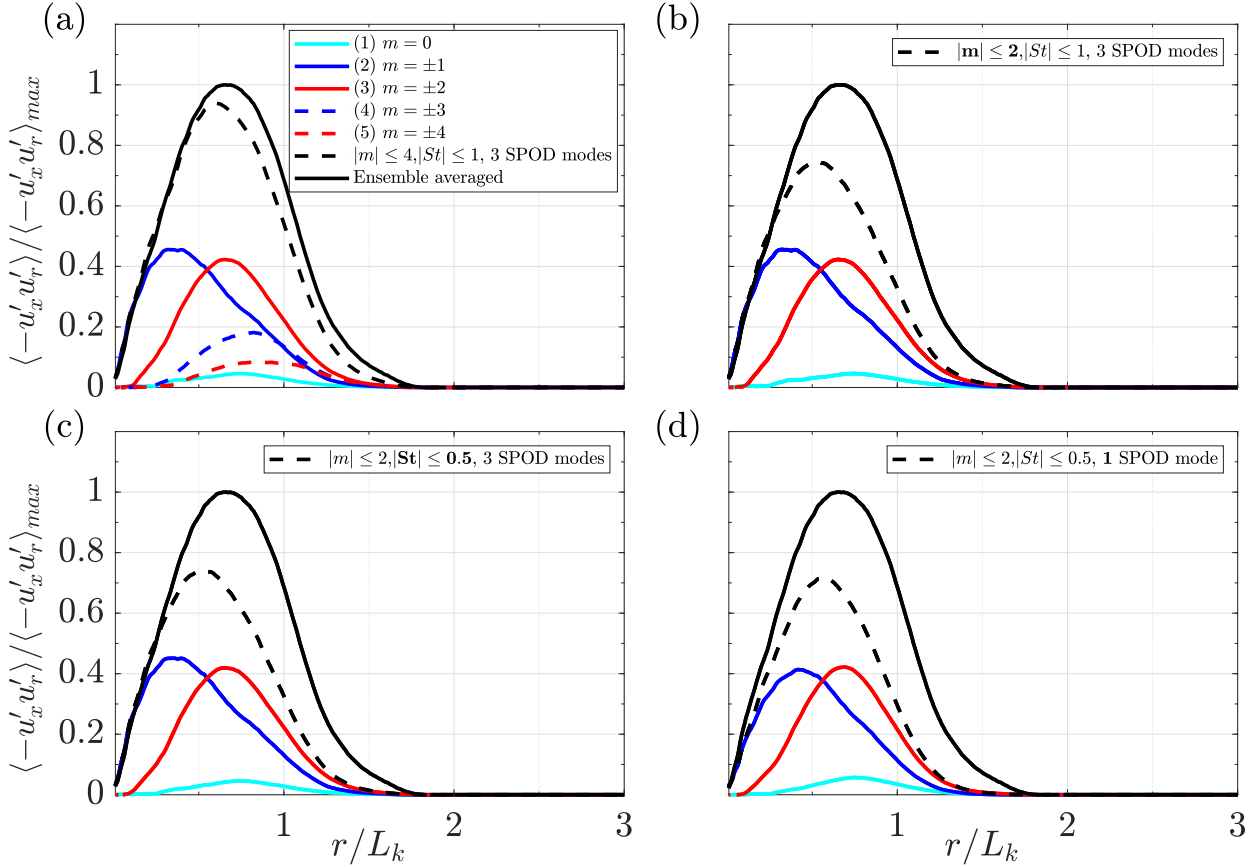}
\caption{Reconstruction of  $\langle u'_{x}u'_{r} \rangle$ at $x/D = 40$ using: (a) $|m| \leq 4$, $|\Str| \leq 1$, $n \leq 3$, (b) $|m| \leq 2$, $|\Str| \leq 1$, $n \leq 3$, (c) $|m| \leq 2$, $|\Str| \leq 0.5$, $n \leq 3$, and (d) $|m| \leq 2$, $|\Str| \leq 0.5$, $n =1$. Radial direction is scaled with TKE-based wake width $L_{k}$ and $\langle u'_{x}u'_{r} \rangle$ is scaled with $\langle -u'_{x}u'_{r} \rangle_{max}$. In each subplot from (b) to (d), one parameter is changed relative to the preceding plot and that parameter is boldfaced in the legend.}
\label{fig:uxur_reconstruction_analysis}
\end{figure}

At $x/D = 40$, the error in reconstruction of $\langle -u'_{x}u'_{r} \rangle_{max}$ is already small, not more than 10 \%. It is found that increasing the azimuthal mode count to $|m| = 10$ is sufficient to obtain complete reconstruction (not shown). It is not necessary to increase the SPOD number beyond $n =3$ and the frequency above $\Str = 1$.  

The  influence of decreasing the number of retained modes on the reconstruction of $\langle -u'_{x}u'_{r} \rangle$ has also been explored.
Figure~\ref{fig:uxur_reconstruction_analysis} illustrates the results for one representative location, $x/D = 40$,  noting that the same procedure at other locations results in the same qualitative conclusions. In Fig. \ref{fig:uxur_reconstruction_analysis}(b), the total number of azimuthal modes included for reconstruction is decreased to $m=2$ keeping $\Str$ and the number of SPOD modes unchanged. The  reconstruction quality is reduced with the reconstructed $\langle u'_{x}u'_{r} \rangle$ capturing around 75\% of the peak $\langle u'_{x}u'_{r} \rangle$ in Fig. \ref{fig:uxur_reconstruction_analysis}(b).  The location of the peak at $r/L_{k} \approx 0.5$ is shifted erroneously closer to the centerline  since the higher-$m$ modes, which peak at $r/L_{k} \approx 0.75$, are now excluded from the reconstruction. An additional decrease in the number of frequencies (Fig. \ref{fig:uxur_reconstruction_analysis} c) does not lead to any noticeable difference  in the reconstruction.  Finally,  the effect of further limiting the modal content  to solely the leading SPOD mode is shown in Fig. \ref{fig:uxur_reconstruction_analysis} (d). Although  the change in the peak of $\langle -u'_{x}u'_{r} \rangle$ is negligible, the
reconstruction  accuracy in 
  in the region $r/L_{k} < 0.25$ suffers.
 Thus, the inclusion of a small number of additional SPOD modes (up to $n =3$ in the present example) beyond $\lambda ^{(1)}$ is needed to accurately capture the trends in $\langle u'_{x}u'_{r}\rangle$ near the centerline. This is somewhat analogous to the TKE reconstruction, where inclusion of more SPOD modes improved the accuracy at the centerline. A notable difference is that, in the case of $\langle u'_{x}u'_{r} \rangle$, it is not necessary to include higher SPOD modes beyond $n=3$ unlike in  the TKE reconstruction where they significantly improved the accuracy.

For high $\Rey$ axisymmetric shear flows, $\langle u'_{x}u'_{r} \rangle$ is the dominant off-diagonal Reynolds  stress.  $\langle u'_{x}u'_{r} \rangle$ extracts energy from the mean flow through turbulent production and transfers it to the TKE.  Reconstructions shown in Fig. \ref{fig:uxur_reconstruction_analysis} indicate that the majority of $\langle u'_{x}u'_{r} \rangle$ is contained in the first few leading SPOD modes of low $m$ and low $\Str$. This is in stark contrast to the reconstruction of TKE. The same set of modes which captured $50\%$ of the centerline TKE, capture about $90\%$ of the peak $\langle -u'_{x}u'_{r}\rangle$. The reconstruction of TKE using varying sets of modes presented in Fig. \ref{fig:k_reconstruction_analysis} showed that the higher modes (modes having large $m$, large $\Str$ and higher SPOD index) can also contribute significantly to TKE when summed together. On the other hand, including more SPOD modes does not significantly improve the Reynolds stress reconstruction. This leads to the conclusion that the interaction between the mean flow and turbulence occurs primarily through more energetic SPOD modes of low $m$ and $\Str$. These modes then transfer the TKE to the less energetic modes as the flow evolves.

\section{SPOD Analysis of locations near the disk}\label{spod_nearbody}

\begin{figure}
	\centering
	\includegraphics[width=\linewidth]{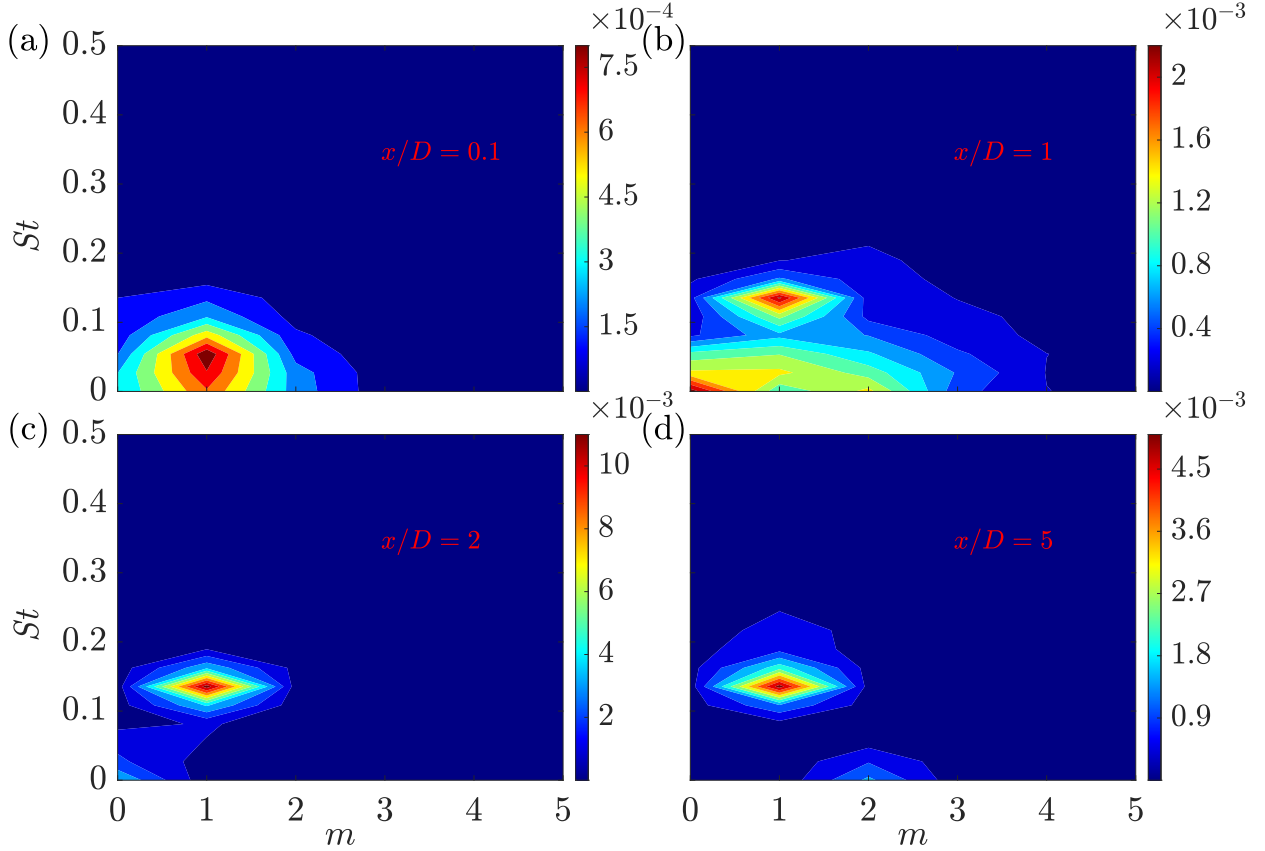}
	\caption{SPOD contour maps showing energy contained in leading SPOD mode, $\lambda^{(1)}$, as a function of azimuthal wavenumber $m$ and frequency $\Str$ at different locations near the disk: (a) $x/D = 0.1$, (b) $x/D = 1$, (c) $x/D = 2$, and (d) $x/D = 5$. The colorbar limits are set according to the maximum values of $\lambda^{(1)}$ over all  ($m$, $\Str$) pairs at the respective $x/D$ locations.}
	\label{fig:contour_maps_eigenvalues_nearbody}
\end{figure}
	
To conclude the results, we present a brief analysis of SPOD spectra at additional locations,  $x/D = 0.1,1, 2, $ and $5$, so as to shed  light on the modal energy distribution close to the disk and its transition to the wake.

Figure \ref{fig:contour_maps_eigenvalues_nearbody} shows the distribution of energy in the $\lambda^{(1)}$ constituent as a function of $m$ and $\Str$ at four streamwise locations: $x/D = 0.1, 1, 2$ and $5$.
 At $x/D = 0.1$ shown in Fig. \ref{fig:contour_maps_eigenvalues_nearbody}(a), most of the energy is concentrated in a narrow range: $\Str \leq 0.1$ and $m \leq 2$. The azimuthal $m=1$ mode shows a distinctive peak at $\Str = 0.054$. In their study of flow past a disk, Berger et al. \cite{berger_coherent_1990} found a spectral peak at $\Str \approx 0.05$ in the axisymmetric mode $m=0$, not $m =1$. They attributed this frequency to  pumping of the recirculation bubble. As will be seen later, the spectral peak at $\Str  \approx 0.05$ does appear in the $m=0$ SPOD eigenspectra at  larger $x/D$. 

The $x/D = 1$ location lies approximately in the middle of the recirculation region and has a complex distribution of energy among  different $m$ and $\Str$. By $x/D = 1$  (Fig. \ref{fig:contour_maps_eigenvalues_nearbody} b), the dominant  $\Str = 0.054$, $m=1$ peak has  disappeared and a new peak at the vortex shedding frequency, $\Str=0.135$, appears. This location marks the initial appearance of the vortex shedding structure which  dominates the SPOD eigenspectra until $x/D \approx 40$ (Fig. \ref{fig:contour_maps_eigenvalues} and \ref{fig:bar_plots_m}). Besides the peak at the VS structure, a major portion of energy in the $\lambda^{(1)}$ constituent is also contained in the region of $\Str \leq 0.05$ distributed over $0 \leq m \leq 3$.  At $x/D = 2$ (Fig. \ref{fig:contour_maps_eigenvalues_nearbody} c), the contour map of $\lambda^{(1)}$ is found to be dominated by the VS structure with $m=1, \Str = 0.135$. The location of $x/D \approx 2$ also marks the end of the turbulent recirculation region behind the disk and the contribution from its high-$m$ modes.
The DH structure ($m=2, \Str \rightarrow 0$) which dominates the far wake  is absent from the energy distribution map of $\lambda^{(1)}$ at $x/D \leq  2$, but  appears as a local peak by $x/D = 5$ (Fig. \ref{fig:contour_maps_eigenvalues_nearbody} d).

\begin{figure}
	\centering
	\includegraphics[width=\linewidth]{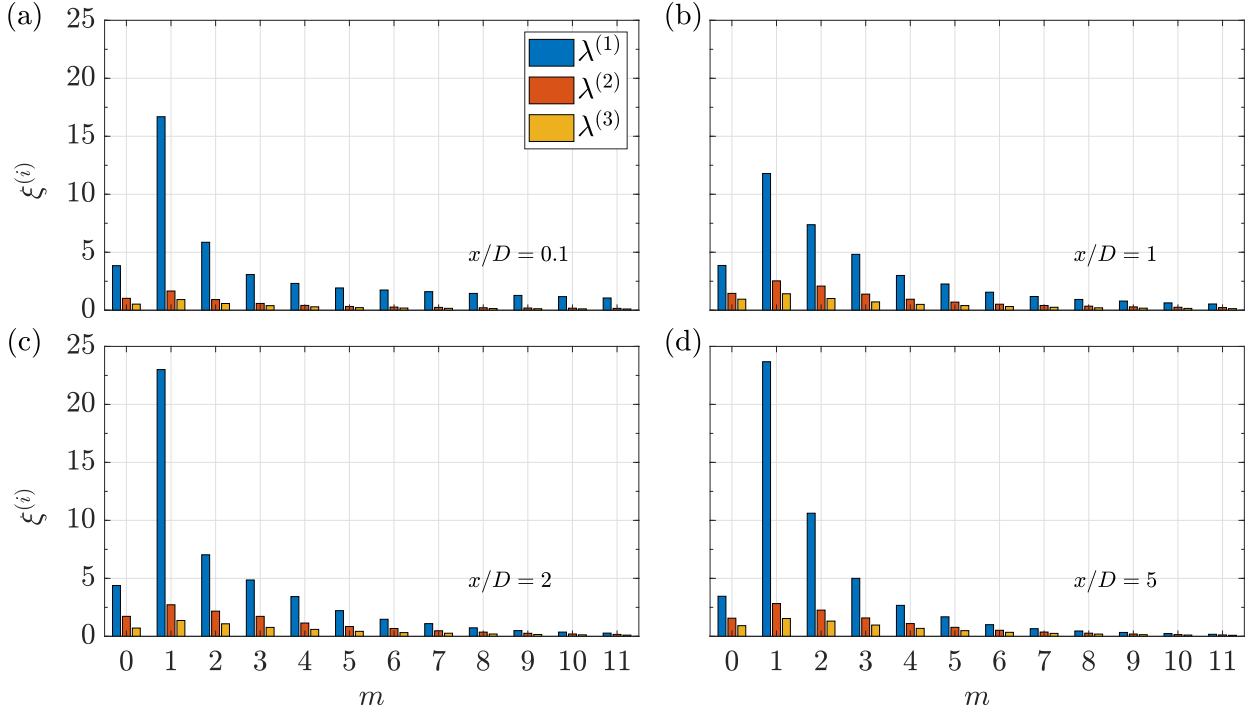}
	\caption{Frequency-integrated eigenspectrum as a function of azimuthal mode number $m$ at different locations in the near wake: (a) $x/D = 0.1$, (b) $x/D = 1$, (c) $x/D = 2$, and (d) $x/D = 5$. Three leading SPOD modes ($\lambda^{(1)}$, $\lambda^{(2)}$, and $\lambda^{(3)}$ ) at each $m$ are shown in terms of their percentage contributions to the area-integrated TKE.} 	
	\label{fig:bar_plots_m_nearbody}.
\end{figure}

The frequency-integrated eigenspectra, at the same four locations of $x/D = 0.1, 1, 2$ and $5$, are shown in Fig. \ref{fig:bar_plots_m_nearbody}, to contrast  the relative importance of different azimuthal modes near the disk. The procedure for obtaining the frequency integrated eigenspectrum is same as that used for Fig. \ref{fig:bar_plots_m}. 
  At all four locations, the $m=1$ mode dominates the frequency-integrated eigenspectrum, followed by $m=2$. At $x/D =2$, the relative energy content in the $m=1$ mode  becomes almost twice  that at 
$x/D = 1$. This sudden increase in the relative importance of the $m=1$ mode is also seen in the energy distribution of $\lambda^{(1)}$ (Fig. \ref{fig:contour_maps_eigenvalues_nearbody}) where the broadband distribution of energy at $x/D = 1$ gives way to a single dominant peak at $m=1, \Str = 0.135$ at $x/D = 2$. By $x/D = 5$, the $m=2$ mode starts gaining relative importance although $m =1$ is still dominant. 
Besides the $m=1$ and $2$ modes,  the major contributors to the frequency-integrated eigenspectra are the $m=0, 3$ and $4$ modes, similar to the previously shown far-wake locations.

\begin{figure}
	\centering
	\includegraphics[width=\linewidth]{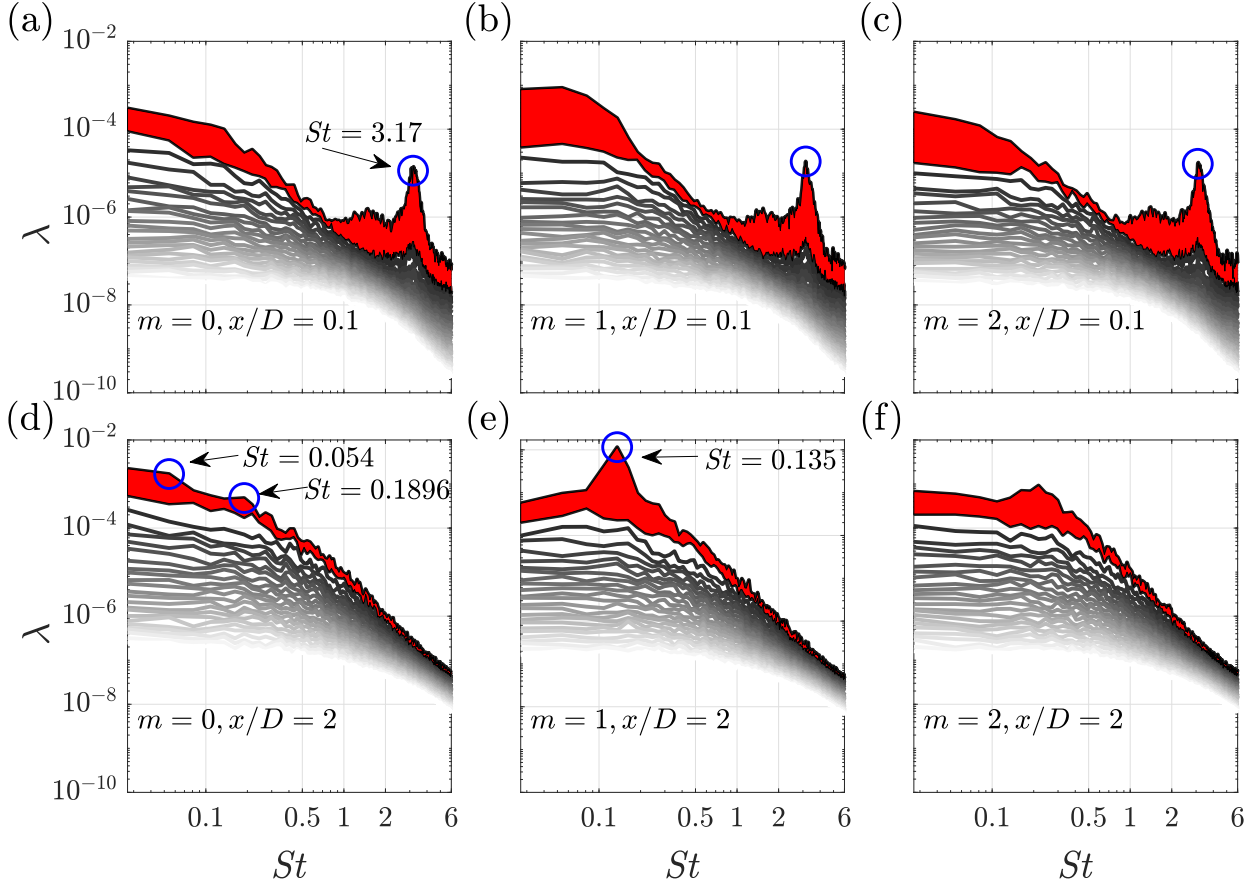}
	\caption{SPOD eigenspectra of $25$ modes (dark to light shade corresponds to high to low energy eigenvalues): (a) $m=0$, $x/D = 0.1$; (b) $m=1$, $x/D = 0.1$; (c) $m=2$, $x/D = 0.1$; (d) $m=0$, $x/D = 2$; (e) $m=1$, $x/D = 2$; (f) $m=2$, $x/D = 2$.} 
	\label{fig:eigenspectra_m012}
\end{figure}

Figure \ref{fig:eigenspectra_m012} shows the SPOD eigenspectra of $m=0, 1$, and $2$ azimuthal modes at two locations: $x/D = 0.1$ (very close to the disk) and $x/D = 2$ (end of the recirculation region). At $x/D = 0.1$ shown in the top row of Fig. \ref{fig:eigenspectra_m012}, a distinct peak at the high frequency of $\Str = 3.17$ is visible in the $\lambda^{(1)}$  eigenspectrum at   all three values of $m$. This high-frequency peak is associated with the shear layer instability in the vicinity of the disk, arising from  the instability of the  boundary layer which separates from the disk. It is also worth mentioning that the SPOD eigenspectra of all three azimuthal modes $m=0, 1, $ and $2$ show a prominent gap between the $\lambda^{(1)}$ and $\lambda^{(2)}$ SPOD modes at and near  the  frequency of $\Str = 3.17$. The study by Berger et al. \cite{berger_coherent_1990} at  $\Rey = 15,000$ identified $\Str = 1.62$ as the frequency related to the shear layer instability, dominated by the $m=0$ mode, and followed by $m=1$ and $2$ modes with equal contributions. 
The  higher $\Str$ identified in the near-disk SPOD eigenspectra here is likely a consequence of the higher $\Rey = 50,000$ of the current study.
Fig. \ref{fig:eigenspectra_m012}(b) shows the eigenspectra of the $m=1$ mode at $x/D = 0.1$.
  At this location, a broad peak around $\Str = 0.05$ is found as was  was also seen in the $\lambda^{(1)}$ contour map at $x/D = 0.1$ in Fig. \ref{fig:contour_maps_eigenvalues_nearbody}(a). 
  
Figure \ref{fig:eigenspectra_m012}(d) shows the SPOD eigenspectra of the $m=0$ mode at $x/D = 2$. At this location, two small peaks can be found  in the $\lambda^{(1)}$  constituent at  $\Str = 0.054$ and  $\Str = 0.189$, respectively. The first of these frequencies is the axisymmetric pumping of the recirculation bubble identified in previous studies \cite{berger_coherent_1990, yang_numerical_2014} of flow past a disk. The  axisymmetric pumping is found to persist, albeit with decreasing strength, until $x/D = 10$.
 The second of these frequencies, $\Str = 0.189$  has not been discussed in the existing literature of flow past a disk. However, a peak at approximately this frequency does exist in the study of Berger et al. \cite{berger_coherent_1990} (see Fig. 12 of their paper) although the authors do not discuss it. This peak at $\Str = 0.189$ in the $m=0$ mode is found to persist for long downstream distances as seen in Fig. \ref{fig:eigenspectra_m034}(a) and (d) indicating that it is a global mode similar to the VS structure.
 
By $x/D = 2$, a distinct peak at $\Str = 0.135$ appears in the $\lambda^{(1)}$ eigenspectrum of  the $m=1$ azimuthal mode (Figure \ref{fig:eigenspectra_m012} e). A large gap between the $\lambda^{(1)}$ and $\lambda^{(2)}$ spectra at  the vortex shedding frequency also appears by $x/D = 2$ implying that the vortex shedding structure dominates the dynamics of the $m=1$ mode from early on. The SPOD eigenspectra of the $m=2$ mode, shown in Fig. \ref{fig:eigenspectra_m012}(f), shows a peak at  $\Str \approx 0.20$. However, it is found that this peak disappears in the eigenspectra of the $m=2$ mode by $x/D = 5$ (not shown here).  

\section{Summary and Conclusions} \label{conclusions}

In the present study, we elucidate the characteristics of the coherent structures in the turbulent wake of a disk at $\Rey  = 50,000$ using the high-resolution LES  database of Chongsiripinyo and Sarkar \cite{chongsiripinyo_decay_2020}. For this purpose, we decompose the flow snapshots into azimuthal modes $m$ followed by  SPOD analysis to further decompose the modes into  nondimensional frequencies $\Str$ and  energy content. The  eigenvalue ($\lambda^{(n)}$) of the SPOD mode $(n)$ of a given $(m,\Str)$ pair  represents the fraction of area-integrated TKE contained in that mode, thus providing an objective framework to rank the SPOD modes based on the energy content. 

SPOD eigenspectra at different streamwise locations beyond $x/D = 10$  show that the  energy in the leading-order SPOD mode ($\lambda^{(1)}$) is predominantly contained in the low azimuthal modes $(m \leq 4)$ and low frequencies of $\Str < 0.4$. At all the streamwise locations, two distinct peaks: (i) $m=1$, $\Str = 0.135$ and (ii) $m=2$, $\Str \rightarrow 0$ are visible in the flow. The first peak has long been established as the vortex shedding (VS) structure in the turbulent wake of a disk.  The importance of the second peak was established by the experimental studies of Johansson et al. \cite{johansson_proper_2002} and Johansson and George \cite{johansson_far_2006-1}. In particular, the double helix (DH) structure with $m=2$ and $\Str \rightarrow 0$ was found to dominate over the VS structure from $x/D = 30$ onward in their study. In the present case, the azimuthal mode $m=2$ emerges as the dominant azimuthal mode at farther downstream distance, beyond $x/D = 60$. This is also the location where the wake defect law transitions from $U_d \propto x^{-0.9}$ to
$U_d \propto x^{-2/3}$, and where the characteristic r.m.s turbulence scales become proportional to $U_d$ during the wake decay. 


When  $\lambda^{(1)}$ corresponding to the azimuthal modes $m=1$ and $m=2$ are scaled by a parameter representing the area-integrated TKE, i.e., $(K_{o}^{1/2}L_{k})^{2}$, it is found that the eigenspectrum of $m=2$  at different streamwise locations beyond $x/D = 20$ collapse perfectly on to a single curve. On the other hand, the scaled eigenspectrum of  $m=1$  at different $x/D$ show a significant spread around the vortex shedding frequency $\Str = 0.135$ while collapsing for higher frequencies $\Str > 0.3$. It is also found that the leading  eigenmode of the DH structure collapses well in the scaled radial coordinates ($r/L_{k}$), unlike the eigenmode of the VS structure. These findings, along with the FFT analysis of the $m=2$ mode, indicate that the DH structure is connected to the local turbulence structure of the flow. On the other hand,  the VS structure is a global mode which originates near the wake generator (further clarified by our near-body SPOD analyses). More studies, preferably with different wake generators, and at further higher $\Rey$ are needed to explore the robustness of the present results regarding the VS and DH modes.

Apart from  $m=1$ and $2$, the summed contribution of the azimuthal modes $m=0, 3,$ and $4$ to the TKE is  also found to be significant. We also characterize  the eigenspectra of these sub-dominant azimuthal modes  in the present work. The SPOD eigenspectra of $m=3$ and $4$ azimuthal modes show features similar to the $m=1$ and $2$ modes, respectively. The SPOD eigenspectra of the $m=0$ mode show a peak at $\Str = 0.189$, in the intermediate ($x/D = 20$) as well as the far wake ($x/D =80$). Further analysis reveals that this spectral peak is present in the SPOD eigenspectra of the $m=0$ mode at near-body locations ($x/D = 2$) too. 
We speculate that the $m=0, \Str = 0.189$ is a global mode (similar to the VS structure) which appears in the wake of the disk at high $\Rey$. Further studies using global resolvent analysis \cite{thomareis_resolvent_2018, yeh_resolvent-analysis-based_2019} may help decipher the physical origin and dynamics of this particular mode. 

Besides analyzing the SPOD eigenspectra and eigenmodes, we also perform reconstruction of TKE and $\langle u'_{x}u'_{r}\rangle$ using SPOD modes at different $x/D$ locations. In the reconstruction of TKE, it is found that the first few modes of each kind (specifically $|m| < 4, |\Str| < 1,$ and $n \leq 3$) captures approximately $50\%$ of  TKE in the central region. The  contributors to the TKE at the centerline ($r = 0$) are the azimuthal modes $m=0$ and $m=1$. Other higher azimuthal modes start off as zero near the centerline and peak in the region,  $0.5 < r/L_{k} < 1$. The three parameters $(m,\Str,n)$ are then systematically varied to test the sensitivity of TKE reconstruction to different parameters. It is found that higher $n$ (more SPOD eigenmodes) is necessary for accurate reconstruction of the centerline TKE, implying that higher SPOD modes are more active near the centerline and are turbulence controlled rather than being associated with coherent motions. Inclusion of additional $m$ and $\Str$ monotonically improves the reconstruction quality of the TKE, considered over the entire wake width. It is worth noting that in the experimental study by Johansson and George~\cite{johansson_far_2006-1} at $\Rey = 26,700$, the first three eigenmodes were able to capture 90\% of the measured streamwise fluctuation energy.  A lower fraction (approximately 50\%) is captured by these modes in the present study  because  small-scale structures contribute more significantly at the higher $\Rey = 50,000$ considered here, and the significantly  higher radial and azimuthal resolution  possible in a simulation-based work enables accounts for the energy carried by  these  structures with small spatial scale.   

As far as the reconstruction of $\langle u'_{x}u'_{r} \rangle$ is concerned, it is worth noting that the primary production term in the TKE equation for a turbulent axisymmetric wake is $-\langle u'_{x}u'_{r}\rangle \partial \langle U\rangle/\partial r$, where $\langle U \rangle $ is the ensemble-averaged streamwise velocity. Thus, $\langle u'_{x}u'_{r}\rangle $ is  responsible for transferring energy from the mean flow to turbulence within the Reynolds-averaged framework. It is found that the same set of modes with low $m$, $\Str$ and $n$, which capture only around 50\% of centerline TKE, almost completely reconstruct the $\langle u'_{x}u'_{r} \rangle$ in the near wake locations.
The implication is that it is only a considerably reduced set of $(m,\Str,n)$ modes
which directly interact with the mean flow. 
Another important result from the reconstruction of $\langle u'_{x}u'_{r} \rangle$ is the dominance of azimuthal modes $m=1$ and $2$, with each capturing different features of the actual profile. The azimuthal mode $m=1$ captures the slope of the actual profiles at $r/D = 0$ while $m=2$ captures the location of peak in the profile. 

The near-body locations, $x/D = 0.1, 1, 2 $ and $5 $ are also investigated using SPOD to characterize the transition of modal content form  the immediate lee of the body to the near wake. The fluctuation energy at these near-body locations is primarily dominated by the $m=1$ mode, which is in turn dominated by the VS mode from $x/D = 1$ onward. In the close proximity of the disk, at $x/D =0.1$, a high frequency peak at $\Str \approx 3.10$ is detected in all three azimuthal modes $m=0, 1, $ and $2$. This frequency value is significantly larger than $\Str = 1.62$ found by Berger et al. \cite{berger_coherent_1990} in their study of coherent structures in the vicinity of the disk. However, the $\Rey$ of the current study is more than three times  the $Re = 15,000$ of the previous study \cite{berger_coherent_1990}. The $\Str \approx 3.1$ peak in the current study is  likely the shear layer instability;  the reason for the comparatively higher magnitude  of $\Str$ is the high $\Rey$ 
of the current study. 

The VS mode appears in the SPOD eigenspectra of $m=1$ at $x/D = 1$ (approximately in the middle of the recirculation region). Surprisingly, the $m=1$ mode shows a peak at $\Str = 0.054$ in the SPOD eigenspectra at $x/D = 0.1$, which then disappears in its SPOD eigenspectra at further downstream locations. This low frequency has been associated to the axisymmetric ($m=0$) pumping of the recirculation bubble in some previous studies \cite{berger_coherent_1990, yang_numerical_2014}. The eigenspectra of $m=0$ mode shows small peaks at $\Str = 0.054$ (related to the pumping of recirculation bubble) and $\Str = 0.189$ at $x/D = 2$. The DH structure which dominates the far wake of the disk starts gaining importance relative to the VS structure at only $x/D = 5$.

\section{Acknowledgments}

We gratefully acknowledge the support of Office of Naval Research Grants N00014-15-1-2718 and N00014-20-1-2253. S.N. would like to thank A. Nekkanti for technical discussions related to the SPOD.  

\bibliography{prf_spod_frinf_references_v2.0}

\end{document}